\documentclass[journal=nalefd,manuscript=letter]{achemso}
\usepackage{graphicx}
\usepackage{dcolumn}
\usepackage{bm}
\usepackage{comment}
\usepackage{epstopdf}
\usepackage{amsmath}
\usepackage{hyperref}
\usepackage{cleveref}
\usepackage{amsmath}
\usepackage{mathrsfs}
\usepackage{mathtools}
\usepackage{physics}
\usepackage{subcaption}
\usepackage{lineno}

\author{A.~C. McRae$^{\S}$, G. Wei$^{\S}$, L. Huang, S. Yigen, V. Tayari,  and A.~R. Champagne}
\email{a.champagne@concordia.ca}
\affiliation{$^{\S}$These two authors made equal contributions. \linebreak Department of Physics, Concordia University, Montr\'{e}al, Qu\'{e}bec, H4B 1R6, Canada}
\title[\texttt{achemso}]
{Mechanical control of quantum transport in graphene}
\begin{document}

\begin{abstract}
Two-dimensional materials (2DMs) are fundamentally electro-mechanical systems. Their environment unavoidably strains them and modifies their quantum transport properties. For instance, a simple uniaxial strain could completely turn off the conductivity of ballistic graphene or switch on/off the superconducting phase of magic-angle bilayer graphene. Here we report measurements of quantum transport in strained graphene which agree quantitatively with models based on mechanically-induced gauge potentials. We mechanically induce \textit{in-situ} a scalar potential, which modifies graphene's work function by up to 25 meV, and vector potentials which suppress the ballistic conductivity of graphene by up to 30 \% and control its quantum interferences. To do so, we developed an experimental platform able to precisely tune both the mechanics and electrostatics of suspended graphene transistors at low-temperature over a broad range of strain (up to 2.6 \%). This work opens many opportunities to experimentally explore quantitative strain effects in 2DM quantum transport and technologies.
\end{abstract}

Because their bulk is also a surface, 2DMs' crystal lattices and electronics are tailored by unavoidable strain fields from their surroundings (substrate, contacts, interfaces, defects). This built-in mechanical tunability (straintronics) offers a wide range of possibilities to optimize quantum technologies (e.g.\ qubits\cite{Alfieri23, Banszerus21, Liu19}, spintronics\cite{Pal23, Wu18, Hanakata18, Molle17}, valleytronics\cite{Li20, Settnes17,Schaibley16,Guinea10}) and many-body quantum phases (e.g.\ superconductivity\cite{Kapfer23, Kim23, Rakib22, Khanjani18}, topological transitions\cite{Kim23, Zhang23, Pantaleon21, Du21, Moulsdale20, Mutch19}, magnetic transitions\cite{Cenker22, Li19, Burch18}). For example, even a simple uniform uniaxial-strain could create high on-off ratio graphene transistors without needing a band-gap \cite{Fogler08,McRae19}, act as topological switch turning a trivial insulator into a quantum spin Hall system \cite{Molle17}, or tune the superconducting phase diagram of magic-angle bilayer graphene \cite{Kazmierczak21}. Strain-engineering experiments on quantum 2DMs have so far focused mostly on non-transport studies\cite{Cenker22, Dirnberger21, Mesple21, Shi20, Li20, Banerjee20, Nigge19, Liu18, Goldsche18, Jiang17, Levy10} which are much less sensitive to long-range strain disorder than transport. While progress has been made in quantum transport experiments\cite{Wang21, Wang20_2, Zhang19, De_Sanctis18, Guan17, Shioya15, Dean10}, a complete control of mechanical strain fields in 2DMs (from substrate, contacts, interfaces, defects) and understanding of their impact on transport has not been achieved. To verify quantitatively theoretical straintronics predictions\cite{Kim23,Miao21}, quantum transport experiments require a precise control of \textit{all} sources of strain over an entire transport device. This experimental challenge, and its accurate modeling, has prevented the experimental verification of even the most canonical quantum transport straintronics prediction. Namely, that a uniaxial strain in graphene can be described accurately by a combination of mechanically-induced scalar and vector gauge potentials mimicking the electromagnetic ones \cite{McRae19, Naumis17, Fogler08}.

We report strain-engineering of quantum transport in graphene which is quantitatively consistent with strain-generated gauge potentials (scalar and vector). First, we developed instrumentation and devices to control accurately all sources of mechanical and electrostatic potentials in ultra-short ($\sim$ 100 nm) suspended graphene channels at low-temperature (1.3 K). Our graphene channels and contacts formed single crystals of uniform width to protect the quantization of their transport modes. We applied a total uniaxial strain up to 2.6 \% and could tune it \textit{in-situ} by over 1 \%. We observed, in four devices, that strain generated a scalar gauge potential, $\phi_{\varepsilon}$, which tuned the work function of graphene by up to 25 meV. We studied in detail two devices, where we could isolate the effects of the vector gauge potentials, $\bm{A_{\text{i}}}$. We observed a reproducible suppression of the ballistic charge conductivity of up to 30 \% as we increased the $\bm{A_{\text{i}}}$. In addition, the vector potentials controlled the phase of quantum transport interferences, acting as a mechanical analog of the Aharonov-Bohm experiment \cite{van_Oudenaarden98}. The experimental data are in good agreement with a theoretical model whose parameters are all extracted directly from measurements and careful data analysis.

\section{Platform for quantitative 2DM quantum transport straintronics}

\begin{figure*}
\includegraphics{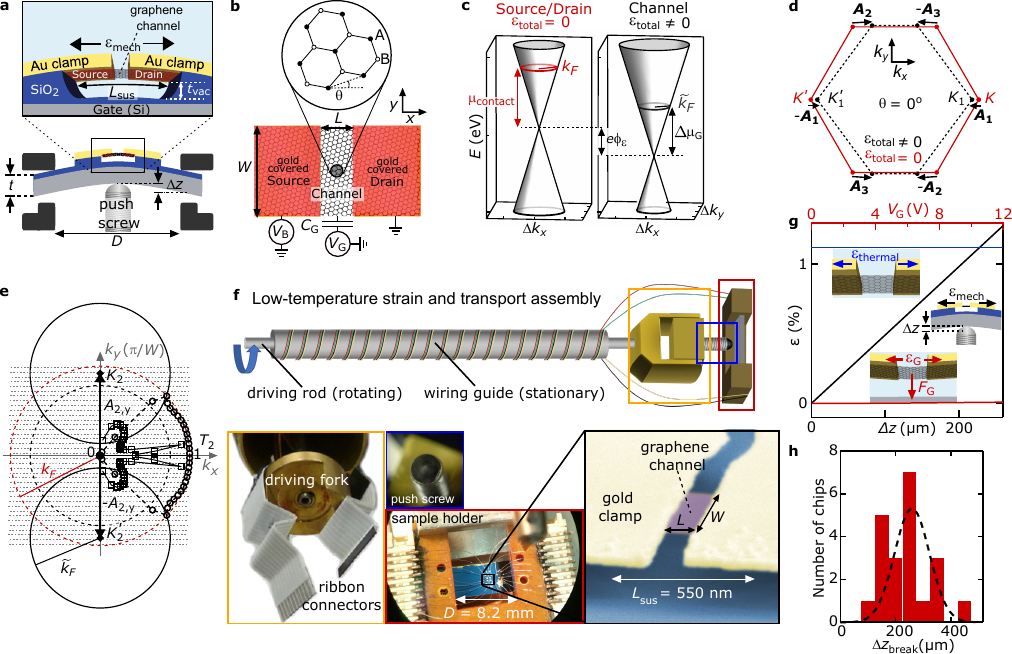}
\begin{flushleft}
\caption{Platform for strain-engineering of quantum transport in 2DMs: applied theory, custom instrumentation, and device design. \textbf{a} Sample holder geometry showing the bending of a Si substrate hosting a suspended graphene transistor. \textbf{b} Top-view diagram of the graphene transistor showing the source, channel, and drain regions. The inset shows the orientation $\theta$ of the crystal lattice with respect to the strain direction. \textbf{c} Band structure in one valley of the graphene's first Brillouin zone (FBZ) in the source/drain (unstrained) and channel (strained). The vertical shift of the bands due to strain is $e \phi_{\varepsilon}$. \textbf{d} FBZ of graphene without strain (red) and with uniaxial strain (black). The displacement of the Dirac points is given by the $\bm{A}_{\text{i}}$ potentials. \textbf{e} The Fermi circles in the $K_{2}$ valley. The strained/unstrained channel's circle are in black solid/dashed lines. The Fermi circle in the unstrained contacts is in red. The open black square and circle markers show the transmission probability, $T_2$, of each mode (subband) when the channel is strained and unstrained, respectively. \textbf{f} Overview of the custom-built instrumentation to transmit mechanical motion inside a cryostat and bend the device's substrate. \textbf{g} Mechanical strain (black trace), $\varepsilon_{\text{mech}}$, and thermal strain (blue trace), $\varepsilon_{\text{thermal}}$, versus $\Delta z$. The gate-induced strain $\varepsilon_{\text{G}}$ is shown in red as a function of $V_{\text{G}}$ (top-axis). \textbf{h} Histogram of the maximum $\Delta z$ achieved before the breaking of Si substrates.} \label{fig1}
\end{flushleft}
\end{figure*}

Figure \ref{fig1} presents the key elements of our experimental platform and applied theoretical model to study quantum transport straintronics in 2D materials. More details can be found in Methods and Supplementary section 1. Figure \ref{fig1}a shows conceptually our mechanical straining method. It consists in bending an ultra-thin Si substrate ($t=$ 200 $\mu$m), over a length $D=$ 8.2 mm, to stretch uniaxially a suspended graphene channel anchored by suspended gold clamps. The contacts and channel are made of a single, uniform width, graphene crystal. Figure \ref{fig1}b shows how the source and drain graphene contacts are covered by gold clamps, which dope them via charge transfer\cite{McRae17, Chaves14, Heinze02}. The overlap area between the gold clamps and each graphene contact is large (few $\mu$m$^2$) and provides slippage-free clamping, as will be shown below. The channel's Fermi energy is controlled via a gate voltage, $V_{\text{G}}$, applied to the Si backplane. The inset shows the crystal lattice and its orientation, $\theta$, which is the angle between the strain direction $x$ and the zig-zag direction of the crystal. This device geometry allows us to accurately model the effect of strain, $\varepsilon_{\text{tot}}$, on the charge carriers\cite{McRae19, Naumis17, Fogler08} by adding gauge potentials to the standard ballistic transport model\cite{Tworzydlo06}.

The mechanically-induced scalar potential, $\phi_{\varepsilon}$, shifts down the Fermi energy in the channel (Fig. \ref{fig1}c, in black) with respect to the unstrained contact regions (in red). Figure \ref{fig1}d shows the unstrained (red) and strained (black) first Brillouin zones (FBZ) for a strain along the zig-zag direction ($\theta =$ 0$^{o}$). The potentials $\bm{A}_{\text{i}}$ generated by $\varepsilon_{\text{tot}}$ shift the momentum positions of the valleys and break the symmetry of the FBZ, such that it becomes necessary to label each valley as $K_{\text{i}}$ and $K'_{\text{i}}$, with $i=$ 1, 2, 3.

Figure \ref{fig1}e shows the effect of $\bm{A}_{\text{2,y}}$ on charge transport in the $K_{\text{2}}$ valley. We note that the $x$-component of the vector potentials has no impact\cite{McRae19, Naumis17, Fogler08}. The  Fermi circles in the contacts (red, radius $k_F$) and the channel (black, radius $\tilde{k_F}$) are shown as dashed lines for $\varepsilon_{\text{tot}}=$ 0 \%. When strain is applied ($\varepsilon_{\text{tot}}=$ 2.61 \% in Fig. \ref{fig2}a), the degeneracy of the $K$/$K'$ Fermi circles in the channel (solid black) is lifted and each valley is shifted up/down by $\pm A_{\text{2,y}}$. The quantized $k_{y}$ = $\pm(n+1/2)\pi/W$ of the conduction modes are shown as dashed horizontal grey lines, and $W$ is the width of the device and $n$ is an integer ranging from 0 up to a maximum set by $k_{F}$. Since $W$ is constant, the total y-momentum is conserved and inside the channel $\tilde{k_{y}}=k_{y}\pm A_{2,y}$. Only the modes whose $k_{y}$ can be matched with an available $\tilde{k_{y}}$ (inside both the black and red circles), have significant transmission amplitude. The calculated transmission probability, $T$, in each conduction mode is shown as open circle (open square) markers for $\varepsilon_{\text{tot}}=$ 0 \% (2.61 \%). The device's charge conductance $G$ is obtained by summing the transmission of all modes in all valleys. We emphasize that it is essential to know the energy scale $\mu_{contact}=\hbar v_F k_F$ to understand quantitatively the impact of the strain-induced potentials $\bm{A}_{\text{i,y}}$'s.

We highlight key features of our strain-engineering quantum transport instrumentation. To create a large, and linear, mechanical force able to bend Si chips (Fig.\ \ref{fig1}a), we designed the assembly shown in Fig.\ \ref{fig1}f and Supplementary Fig.\ S2. A stainless-steel rod transmits mechanical motion to a driving fork (gold-framed inset) which rotates a fine-threaded screw (blue-framed inset) pushing the back of the device's substrate (red-framed inset). The black-framed inset shows a tilted scanning electron microscope (SEM) image of one of our devices, whose lengths are $L\approx$ 100 nm, width $W\approx$ 1 $\mu$m, and overall suspension length $L_{\text{sus}}\approx$ 500 nm. Figure \ref{fig1}g displays the three main sources of strain acting on the suspended graphene channels. The mechanically-tunable strain $\varepsilon_{\text{mech}}$ based on device dimensions (black trace) is shown as a function of the vertical push screw displacement $\Delta z$. The thermal contraction/expansion induced strain $\varepsilon_{\text{thermal}}$ (blue trace) is constant at fixed temperature and around 1\% in our devices (Supplementary section 3). The gate-voltage induced strain $\varepsilon_{\text{G}}$ (red trace) is negligible due to the very short length of the channel\cite{McRae19}. The geometrically predicted range for $\varepsilon_{\text{mech}}= \Delta x/L= (3L_{\text{sus}}t/D^2)\Delta z/L$ is set by the maximum substrate movement $\Delta z_{\text{break}}\approx$ 260 $\mu$m before it fails, as shown in Fig. \ref{fig1}h. A crucial feature of the instrumentation is that it permits to \textit{independently} control $\mu_{\text{channel}}$, with $V_{\text{G}}$, and the $\bm{A}_{\text{i,y}}$ which are linearly proportional to $\varepsilon_{\text{tot}} = \varepsilon_{\text{mech}} + \varepsilon_{\text{thermal}}$.

\section{Calibration of the mechanical and thermal strains}

\begin{figure*}
\includegraphics{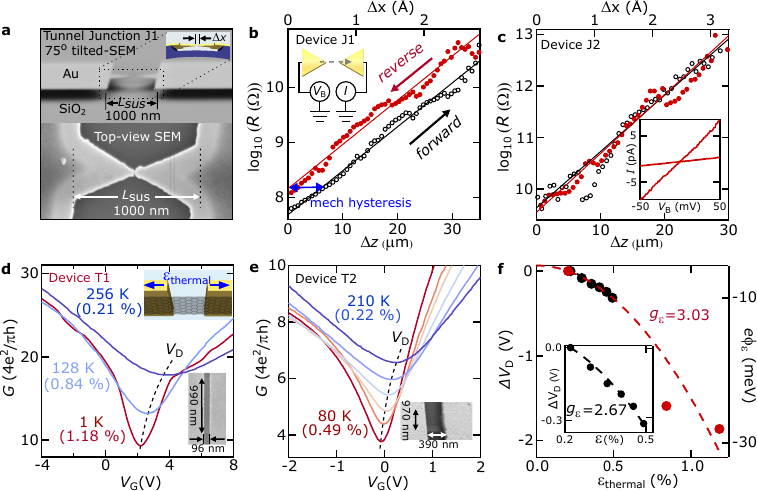} \begin{flushleft}
\caption{Mechanical and thermally-induced strain calibration}\textbf{a} Tilted-SEM (top) and top-view SEM (bottom) images of the tunnel junction Device J1. \textbf{b} Log$_{10}(R)$ vs $\Delta z$ for Device J1. The inset shows the schematics of the measurement circuit. \textbf{c} Log$_{10}(R)$ vs $\Delta z$ for Device J2. The inset shows individual $I$ - $V_{\text{B}}$ traces, whose inverse slope give the data in the main panel. \textbf{d} $G$ - $V_{\text{G}}$ data in Device T1 at temperatures of 256 K, 128 K, and 1 K. The lower inset shows the device's dimensions. \textbf{e} $G$ - $V_{\text{G}}$ data in Device T2 at temperatures ranging from 210 K to 80 K. The inset shows the device dimensions. \textbf{f} The relative Dirac point shift, $\Delta V_{\text{D}}$, versus $\varepsilon_{\text{thermal}}$ for Devices T1 (red data) and T2 (black data). The dashed line is a theoretical fit. The inset shows a zoom-in of the T2 data. \label{fig2}
\end{flushleft}
\end{figure*}

To quantify the gauge potentials in our devices, we carefully calibrated both $\varepsilon_{\text{mech}}$ and $\varepsilon_{\text{thermal}}$. The mechanically-tunable strain comes from the lateral displacement $\Delta x$ of the gold clamps in Fig.\ 1a. This displacement can be calibrated by shaping the gold clamps (without graphene) into a bow-tie bridge, which is electromigrated \cite{Champagne05} to create a tunnel junction between two gold tips (Fig. \ref{fig2}a). The tunnel current is exponentially suppressed by the width of the vacuum gap, whose length is modulated mechanically with $\Delta x$. Figures \ref{fig2}b-c show the resistance, $R$, vs. $\Delta x$ (top axis) and  $\Delta z$  (bottom axis) for two tunnel junctions (Devices J1 and J2). The lower inset of Fig.\ \ref{fig2}c shows two examples of the raw current data, $I$, versus bias voltage, $V_{\text{B}}$, measured at each mechanical position. Their inverse slope $R$ is plotted in the main panel. The $R$ vs. $\Delta z$ data are linear on a log scale, extremely stable over time, and reproducible over multiple back-and-forth mechanical sweeps. As expected there is a small mechanical hysteresis (Fig.\ 2b) stemming from the torsion of the driving rod. This hysteresis is reproducible and is systematically removed from data sets presented thereafter (as shown in Fig. 2c). Combining the data sets, we extract a calibration of $\Delta x /\Delta z = $ 9.0 $\pm$ 1 $\times$ 10 $^{-6}$ in very close agreement with the geometrically predicted  $\Delta x /\Delta z$ = 9.0$\times$ 10$^{-6}$. The precision and stability of $\Delta x$ is $\sim$ 5 pm.

Figure \ref{fig2}d-e show $G$ - $V_{\text{G}}$ data at various temperatures from two samples (Devices T1 and T2) used to calibrate $\varepsilon_{\text{thermal}}$ in our devices. The channel dimensions $L$ and $W$ are shown in the insets and $L_{\text{sus}}=$ 550 and 1390 nm, respectively. We used the thermal contraction/expansion coefficients of gold and graphene to calculate $\varepsilon_{\text{thermal}}$ at each temperature (Supplementary section 3). We then verified experimentally these calculations using graphene's linear work-function shift with strain. This is due to the strain-dependent scalar potential\cite{Choi10} $\phi_{\varepsilon}= (g_{\varepsilon}/e)(1-\nu)\varepsilon_{\text{total}}$, where $\nu$ is the Poisson ratio 0.165 and $g_{\varepsilon}\approx 3.0$~eV. This $\phi_{\varepsilon}$ creates a shift of the $G$ - $V_{\text{G}}$ minimum's position, $\Delta V_{\text{D}}$, given by,
\begin{equation}\label{Eq1}
\Delta V_{\text{D}}= -\frac{e}{c_G}\frac{g^2_\varepsilon}{\pi(\hbar v_F)^2}(1-\nu)^2\varepsilon\textsubscript{total}^2
\end{equation}
where $c_G$ is the capacitance per unit area. In Fig.\ \ref{fig2}f, we plot $\Delta V_{\text{D}}$ vs. $\varepsilon_{\text{thermal}}$ for Devices T1 (red markers) and T2 (black markers) and fit them with Eq. \ref{Eq1} to extract $g_{\varepsilon}$ = 3.03 and 2.67, in good agreement with theory \cite{Grassano20, Choi10}. The inset of Fig.\ \ref{fig2}f shows a zoom-in of the data and fit for Device T2. This confirms that we understand with a good accuracy $\varepsilon_{\text{thermal}}$ in our devices. The dominant source of uncertainty comes from the measurement error on $L_{\text{sus}}$, $\pm$ 50 nm, and leads to a systematic uncertainty in the extracted $\varepsilon_{\text{thermal}}$ of about one part in ten. We now present transport data as a function $\varepsilon_{\text{mech}}$.

\section{Mechanical tuning of graphene's work function}

\begin{figure*}
\includegraphics {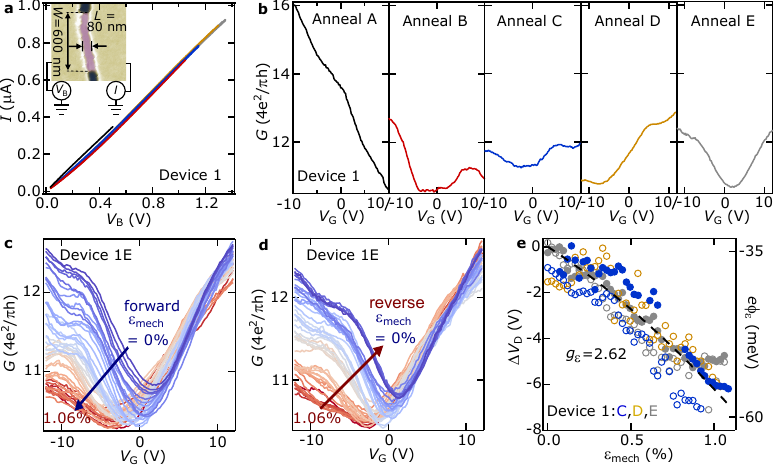}
\begin{flushleft}
\caption{Mechanically-tunable scalar potential and work function in graphene.}\textbf{a} $I$ - $V_{\text{B}}$ data showing successive Joule annealing steps (A-black, B-red, C-blue, D-gold, E-grey) in Device 1. The inset shows the dimensions of Device 1. \textbf{b} $G$ - $V_{\text{G}}$ data in Device 1 after each of the annealing steps in \textbf{a}. \textbf{c-d} $G$ - $V_{\text{G}}$ data in Device 1E, for the forward and reverse $\varepsilon_{\text{mech}}$ sweeps ranging from 0 \% (dark blue trace) to 1.06 \% (dark red trace). Each data trace is equally spaced apart in $\varepsilon_{\text{mech}}$. \textbf{e} The relative Dirac point shift $\Delta V_{\text{D}}$ (left axis) and  $e\phi_{\varepsilon}$ (right axis) versus $\varepsilon_{\text{mech}}$ in Devices 1C (blue data), 1D (gold data), and 1E (grey data). The open/solid markers are from the forward/reverse $\varepsilon_{\text{mech}}$ sweeps. The dashed black line is a theoretical fit. \label{fig3}
\end{flushleft}
\end{figure*}

We used $\varepsilon_{\text{mech}}$ to apply tunable gauge fields to two devices, Device 1 (2)  has dimensions $L=$ 80 (100) nm, $W=$ 600 (850) nm, and $L_{\text{sus}}=$ 550 (570) nm as visible in the inset of Fig.\ \ref{fig3}a and Fig.\ S4a (Fig.\ S5). Before studying our devices, we used Joule annealing \cite{Yigen14, Bolotin08} to reduce the density of randomly fluctuating charge dopants from impurities, $n_{rms}$, and to modify $\mu_{contact}$. Figure \ref{fig3}a shows $I$ - $V_{\text{B}}$ data for successive annealing steps (A-black, B-red, C-blue, D-gold, E-grey) in Device 1, for more details see Methods. The corresponding $I$ - $V_{\text{G}}$ data after each annealing step are shown in Figure \ref{fig3}b. The inset of Fig. \ref{fig3}a shows that the gold clamps' edges have a slight angle with respect to the channel, $\approx$ 10$^o$, which we found to make our devices more resilient to Joule annealing. This geometry was found in previous work to only change the transmission and conductance by $\approx$ 2 \%\cite{Low09}. Since the uncertainties on the dimensions of our devices are around 10 \%, we neglect this much smaller correction in our analysis.

Figure 3b shows that the width of the $I$ - $V_{\text{G}}$ conductance minimum is reduced by the successive annealing steps, meaning that $n_{rms}$ decreases. It is well known that annealing gold thin-films reduces their oxygen content, modifies their work function \cite{Chaves14, Heinze02}, and thus their charge transfer to the underlying graphene contacts in our Devices. A careful observation of Fig.\ \ref{fig3}b reveals that the $I$ - $V_{\text{G}}$ data asymmetry changes from Anneal B to E to show that the graphene contacts evolve from being $p$-doped to having a minimal doping (Anneal C) to being $n$-doped (Anneal E).

We show in Figs.\ \ref{fig3}c-d the $G$ - $V_{\text{G}}$ raw data measured in Device 1E over the forward and reverse $\varepsilon_{\text{mech}}$ sweep from 0 \% (dark blue) to 1.06 \% (dark red). We see a similarly smooth and reversible progression of data at all anneal configurations studied (Figs.\ S4-S8). The reversibility of the mechanical sweeps indicates clearly that the gold-graphene clamps are able to hold without slippage the channel up to the maximum strain applied, $\varepsilon_{\text{total}} = \varepsilon_{\text{thermal}} + \varepsilon_{\text{mech}} =$ 1.55 \% + 1.06 \% = 2.61 \%. The $\Delta V_{\text{D}}$ of each trace in Fig.\ 3c(d) are shown in Fig.\ \ref{fig3}e as open (solid) grey circles. The $\Delta V_{\text{D}}$'s for Devices 1D and 1C (Fig.\ S4) are shown as open gold circles and open (solid) blue circles. The right-hand side $y$-axis of Fig.\ \ref{fig3}e shows the corresponding shift in graphene's work function, $e\phi_{\varepsilon}$. It was tuned \textit{in-situ} by 25 meV and reached a maximum shift of 55 meV, much larger values than in previous work \cite{Wang21}. Fitting the data with Eq.\ \ref{Eq1}, we extract $g_{\varepsilon}=$ 2.62 which is in agreement with both theory \cite{Grassano20, Choi10} and the values extracted in Fig.\ \ref{fig2}f. Data showing the effect of the mechanically-tunable $\phi_{\varepsilon}$ in Device 2 are shown in Fig.\ S5, and give $g_{\varepsilon}=$ 2.72. Collectively, the data in Figs.\ 2-3 and S4-S5 show a precise mechanical control of $\phi_{\varepsilon}$ and work function engineering in graphene. The quantitative experimental understanding of this ``mechanical-gating" effect has far reaching impact on 2DM research and applications, since any crystal deposited on any substrate experiences strains. We now turn our attention to the mechanically-generated $\bm{A_{\text{i}}}$ vector gauge potentials.

\section{Mechanical suppression of graphene's ballistic conductivity}

\begin{figure*}
\includegraphics {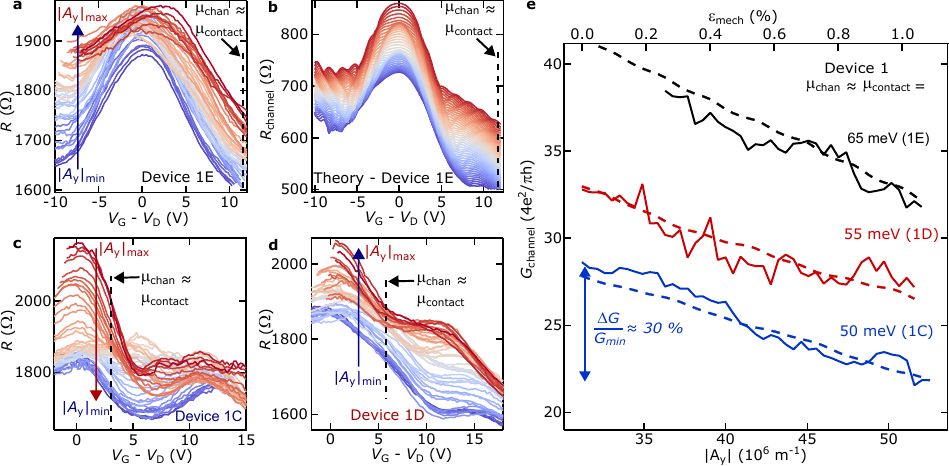}
\begin{flushleft}
\caption{Mechanically-tunable conductance and vector potentials in quantum transport} \textbf{a} $R$-($V_{\text{G}}-V_{\text{D}}$) data in Device 1E for the forward $\varepsilon_{\text{mech}}$ sweep ranging from 0 \% (dark blue trace) to 1.06 \% (dark red trace). \textbf{b} Theoretical calculation of the data in \textbf{a}. \textbf{c} $R$-($V_{\text{G}}-V_{\text{D}}$) data in Device 1C for the reverse $\varepsilon_{\text{mech}}$ sweep ranging from 1.06 \% (dark red trace) to 0 \% (dark blue trace).  \textbf{d} $R$-($V_{\text{G}}-V_{\text{D}}$) data in Device 1D for the forward $\varepsilon_{\text{mech}}$ sweep ranging from 0 \% (dark blue trace) to 1.06 \% (dark red trace). \textbf{e} $G_{\text{channel}}$-$|A_{y}|$ for data extracted along the vertical dashed lines in \textbf{a, c, d} for Devices 1E, 1D, and 1C, respectively. \label{fig4}
\end{flushleft}
\end{figure*}

To isolate the effect of the mechanically-induced gauge potentials $A_{\text{i,y}}$ on our transport data, we removed the $\Delta V_{D}$ shifts from the raw data. Figure \ref{fig4}a shows the resulting $R$ - ($V_{\text{G}}-V_{\text{D}}$) for $\varepsilon_{\text{mech}}$ from 0 \% (dark blue) to 1.06 \% (dark red) in Device 1E. We see that $R$ changes smoothly over the full mechanical range, and that the data are reproducible as shown in Fig. S6a for the reverse sweep. To accurately model how the $A_{\text{i,y}}$ modify $R$, we need to extract the device parameters: $\mu_{contact}$, $\theta$, and $n_{rms}$. This procedure is detailed in Supplementary section 6 and summarized here. An estimate for $n_{rms} \approx 1.5\times 10^{11}$ cm$^{-2}$ (Device 1E) can be extracted from the half-width at half-maximum in Fig. \ref{fig4}a (at mid-strain, $\varepsilon_{\text{mech}}=$ 0.53 \%), and is confirmed by a detailed comparison with calculations (Fig. S6). We extracted $\theta$ by following previous theoretical work\cite{Fogler08} which showed that when both $k_{F}$ and $\tilde{k_{F}}$ are larger than all $|A_{\text{i,y}}|$, $R$ is given by Eq.\ \ref{Eq2} and does not depend significantly on $\mu_{contact}$. These conditions are achieved in Device 1E when ($V_{\text{G}}-V_{\text{D}}$) $\geq$ 10 V. For instance, $R$ along the vertical dashed line in Fig. \ref{fig4}a is given by
\begin{equation}\label{Eq2}
R \approx\frac{h}{e^2W}(\frac{4}{\pi}\tilde{k_{F}} - |A_{\text{y}}(\theta)|)^{-1}
\end{equation}
where $|A_{\text{y}}(\theta)|$ is the average of the three $|A_{\text{i,y}}(\theta)| \propto \varepsilon_{\text{total}}$ (see Methods).

The extracted vertical $\Delta R$ - $|A_{\text{y}}(\theta)|$ data varies linearly and its slope gives $\theta$ = 2.0$^{o}$ $\pm$ 0.5$^{o}$ in Device 1 (see Fig. S6b). Focusing on the slope of the experimental $\Delta R$ - $\varepsilon_{\text{mech}}$ data removes the effect of the series resistance, $R_{s}$, arising from the injection/extraction of the carriers between the gold film and source/drain graphene contacts. Finally, to extract the parameter $\mu_{contact}$, we used the fact that it controls the magnitude of $\Delta R$ in both $\Delta R$ - ($V_{\text{G}}-V_{\text{D}}$) and $\Delta R$ - $\varepsilon_{\text{mech}}$ data. We compared systematically data and theory for various $\mu_{contact}$, and found $\mu_{contact}$ = 65 $\pm$ 5, 55 $\pm$ 5, and 50 $\pm$ 5 meV in Devices 1E, 1D, and 1C respectively (see Fig. S6). We then calculated theoretically, without any free parameter, $R_{channel}$ - ($V_{\text{G}}-V_{\text{D}}$) for Device 1E as shown in Fig. 4b. The $R$ - ($V_{\text{G}}-V_{\text{D}}$) experimental data for Devices 1C and 1D are in Figs. \ref{fig4}c-d, and the corresponding calculations are in Figs. S6f,j. The rigid vertical offset between Device 1 data (e.g.\ Fig.\ \ref{fig4}a) and theory (e.g.\ Fig.\ \ref{fig4}b) is understood as $R_{\text{channel}}=R-R_{s}$, with $R_{s}$ = 1100 $\pm$ 60 $\Omega$.

We observe the quantitative agreement of a wide range of data from Devices 1C, 1D, and 1E with theory, and a good qualitative agreement in Device 2 (Fig. S7) whose $n_{rms}$ is much larger $\approx$ 4.0 $\times$ 10$^{11}$ cm$^{-2}$. In Fig. 4e we show the measured (calculated) $G_{channel}$ in Devices 1C, 1D, 1E as solid (dashed) traces. To meaningfully compare data from different anneals, we focus on data for which $\mu_{channel}$ = $\mu_{contact}$ from each data set. We see in Fig.\ \ref{fig4}e that $G_{channel}$ decreases almost linearly with increasing $|A_{\text{y}}|$. The relative decrease, $\Delta G / G_{\text{min}}$, with $|A_{\text{y}}|$ is largest for the smallest $\mu_{contact}$, and reaches up to $30$ \% in Device 1C. This can be understood based on Fig.\ \ref{fig1}e. We see that shrinking the red circle, $k_{F} = \mu_{contact}/(\hbar v_{F})$, increases the proportion of modes energetically forbidden for a given $|A_{\text{y}}|$. The data and calculations in Figs.\ \ref{fig4}, S6-S7 demonstrate than we can reproducibly decrease the ballistic conductance, and create uniform gauge vector potentials, in graphene using mechanical strain. Beyond the effect of strain on the magnitude of conductance, we are also interested in its effect on the quantum phase of ballistic charge carriers.

\section{Mechanical tuning of quantum transport interferences}

\begin{figure*}
\includegraphics {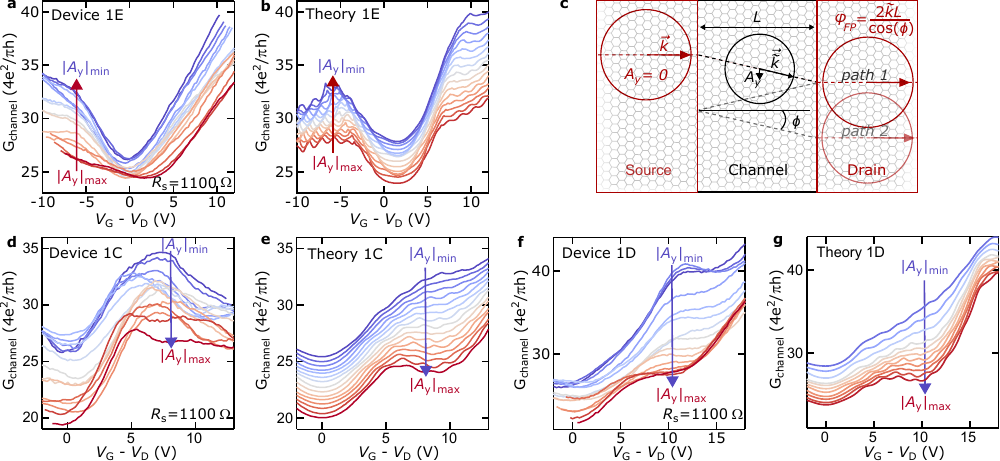}
\begin{flushleft}
\caption{Mechanically-tunable quantum interferences}\textbf{a} $G_{\text{channel}}$ - ($V_{\text{G}}-V_{\text{D}}$) data for a reverse $\varepsilon_{\text{mech}}$ sweep ranging from 1.04 \% (dark red trace) to 0 \% (dark blue trace) in Device 1E. The vertical red arrow shows the location of a strain-tunable quantum interference.  \textbf{b} Theoretical modeling of the data in \textbf{a}. \textbf{c} Diagram showing the interfering paths of ballistic transport in one conduction mode (subband) when a strain-generated $A_{i,y}$ is present. \textbf{d}  $G_{\text{channel}}$ - ($V_{\text{G}}-V_{\text{D}}$) data for $\varepsilon_{\text{mech}}$ sweep ranging from 0 \% (dark blue trace) to 1.04 \% (dark red trace) in Device 1C. The vertical blue arrow shows the location of a strain tunable quantum interference. \textbf{e} Theoretical modeling of the data in \textbf{d}. \textbf{f} $G_{\text{channel}}$ - ($V_{\text{G}}-V_{\text{D}}$) data for $\varepsilon_{\text{mech}}$ sweep ranging from  0 \% (dark blue trace) to 1.04 \% (dark red trace) in Device 1D. The vertical arrow shows the location of a strain tunable quantum interference. \textbf{g} Theoretical modeling of the data in \textbf{f}.\label{fig5}
\end{flushleft}
\end{figure*}

We can use quantum transport interferences to measure strain-induced phase shifts in the wavefunction of charge carriers. Figures \ref{fig5}a-b show the $G_{\text{channel}}$ - ($V_{\text{G}}-V_{\text{D}}$) data and calculations for Device 1E, when $|A_{\text{y}}| = (|A_{\text{y,1}}|+|A_{\text{y,2}}|+|A_{\text{y,3}}|)/3$ is decreased from 5.25 to 3.12 $\times$  10$^{7}$ m$^{-1}$. To improve data clarity, each data trace in Figs.\ \ref{fig5} and S8, were generated by averaging three consecutive traces from the corresponding raw data in Figs. \ref{fig4} and S6. In Figs. \ref{fig5}a-b and S8a, we observe a broad interference maximum in $G$ data around $V_{\text{G}}-V_{\text{D}} \approx$ -6 V, and see that both its magnitude and shape change smoothly as a function of strain. Data and modeling for Devices 1C, 1D, and Device 2 show the mechanical tuning of similar quantum interferences located along the vertical arrows in Figs. \ref{fig5}d-e, \ref{fig5}f-g and Fig. S8e-f, respectively.

The change in $G_{channel}$'s magnitude was discussed in Fig.\ \ref{fig4} and is due to the drop in transmission $T$ as $|A_{\text{y}}|$ increases the carrier's average trajectory angle as shown in Fig.\ \ref{fig5}c. The more subtle change in the line shape of the $G_{channel}$ - ($V_{\text{G}}-V_{\text{D}}$) resonance is explained conceptually in Fig.\ \ref{fig5}c as the superposition of the paths of transmitted and reflected charge carriers leading to quantum (Fabry-P\'{e}rot) interferences.

For example, Fig.\ \ref{fig5}c shows how $|A_{\text{i,y}}|$ changes the carrier trajectory in one mode by an angle $\phi$. When carriers reach the interface between the channel and drain region, they can either be transmitted (path 1, black) or reflected (path 2, grey). The respective lengths of paths 1 and 2 are $L/\cos(\phi)$ and 3$L/\cos(\phi)$. The total transmission amplitude for this mode is the sum of the two paths' amplitudes (neglecting much smaller higher-order terms), and depends on the phase difference $\varphi_{FP, A_{i,y}} = 2\tilde{k_{F}}L/[\cos(\phi(|A_{i,y}|))]$. Therefore, we can tune mechanically this Fabry-P\'{e}rot phase, and the resulting shape of the $G$ resonance, using a gauge potential as in an Aharonov-Bohm experiment\cite{van_Oudenaarden98}. This effect is seen in the data and calculations of all Devices (Figs.\ \ref{fig5} and S8), and there is a good qualitative agreement between the data and theory. The quantitative discrepancies are expected given that we avoided fine tuning our model by using a constant value $L$ for the electrostatic length of the device (from SEM imaging). However, $L$ is expected to be shorten ($\sim$ 10 nm) by $V_{\text{G}}$-dependent electrostatic barriers forming at the channel-contact interfaces \cite{Laitinen16}.

We can get an order of magnitude estimate for the Aharonov-Bohm-like phase introduced using $\varphi_{AB}= \varphi_{FP,A_{\text{max}}} - \varphi_{FP,A_{\text{min}}}$. At the location of the vertical arrow in Fig. \ref{fig5}a, $V_{\text{G}}-V_{\text{D}} =$ -5.75 V, $\tilde{k_F}$= 7.94 $\times$ 10$^{7}$ m$^{-1}$, and  $|A_{\text{2,y}}|$ = 9.02 to 5.35 $\times$ 10$^{7}$ m$^{-1}$. Using $L=$ 80 nm and focusing on the mode for which $\phi \approx$ 0 at the minimum $|A_{\text{2,y}}|$, we obtain $\varphi_{AB}\approx$ $\pi$/2 for the phase difference between the first (dark red) and last (dark blue) trace. This confirms that we can modulate the phase of the quantum interferences very significantly, and matches qualitatively with the experimental data and full calculations in Figs.\ 5 and S8. The data provide compelling evidence that we can \textit{mechanically} tune the quantum phase of charge carriers and attribute this effect to vector gauge potentials.

\section{Conclusions}

In summary, we developed an experimental platform permitting a precise control of uniaxial strain in 2DMs during quantum transport measurements at low temperature. We removed substrate-induced strains by suspending the channel, and then controlled and quantified the effects of thermally-induced and mechanically-tunable strain on quantum transport in graphene. We showed that strain can modify the work function of graphene by over 25 meV, making this effect relevant even for room-temperature 2DM-based quantum technologies\cite{Pal23}. We observed that a $\approx$ 1 \% \textit{in-situ} unaxial strain in graphene creates a smooth and reproducible suppression of the ballistic charge conductance of up to 30 \%. The magnitude of this conductance suppression increased as we reduced the Fermi level in the contact electrodes, showing the importance of contact engineering to understand quantitatively quantum straintronics devices \cite{Kim23}. The reported data matched very well with a theory \cite{McRae19, Naumis17, Fogler08} based on mechanically-induced gauge potentials ($\phi_{\varepsilon}, A_{\text{i,y}}$) analogous to electrostatic ones. Finally, we observed mechanically-tunable quantum transport interferences. We provided a simple interpretation of their shape based on the $A_{\text{i,y}}$ potentials, using an analogy to the Aharonov-Bohm experiment, and found a good qualitative agreement between the data and full calculations.

We expect that our work will open many opportunities to control strain fields with a precision level suited for quantitative quantum transport studies in 2DMs and heterostructures. For instance, our device design allows to clamp mechanically 2DMs from their top surface only. This will permit to apply uniaxial heterostrain in twistronics systems\cite{Kapfer23,Kogl23,Gao23} and magic-angle graphene \cite{Balents20} to explore many-body quantum straintronics\cite{Mesple21, Pantaleon21, Moulsdale20, Wu19}. Moreover, a quantitative control of quantum electro-mechanical transport in 2DMs will be essential to harness their full potential for quantum technologies\cite{Alfieri23,Liu19}.
\newline

\textbf{Methods} \\

\textbf{Conductivity measurements}
Resistance, $R= V_{B}/I$, and conductance, $G = I/V_{B}$, measurements were done using the circuit in Fig. \ref{fig1}b at a temperature of 1.3 Kelvin, unless specified otherwise. The bias voltage used was 0.50 mV for most data, except for Device 2 data where $V_{B}=$ 1.00 mV.

\textbf{Instrumentation}
The low-temperature quantum transport strain instrumentation is shown in Fig. 1f and S2. It is based on a modified a top-loading probe insert for a IceOxford He-3 cryostat, on which we added a carefully designed and tested mechanical assembly. A stepper motor and 100:1 reduction gearbox drive a stainless steel rod which transmit mechanical motion from room temperature to inside the cryostat via a vacuum flange. The transmission rod is terminated with a u-shape fork which rotates a 100 thread/inch push screw on the back side of the sample's substrate to bend it with very fine resolution. The stepper motor is always powered off during transport measurements.

\textbf{Sample nanofabrication} We use e-beam lithography (PMMA/MMA bilayer) to define the gold clamps on the graphene crystals. Kish graphite was used to mechanically exfoliate graphene on a SiO$_\mathrm{{2}}$(300 nm)/Si substrate. The thickness of the crystals was verified via Raman spectroscopy. The mechanically-tunable devices (J1, J2, Device 1 and Device 2) were fabricated on 200 $\mu$m-thick wafers, while the thermally-tunable devices (Devices T1 and T2) were fabricated on 500 $\mu$m-thick substrates. We evaporated a 100-nm thick gold film (no adhesion layer) to define the gold clamps. We used a wet buffered HF oxide etch (BOE) to freely suspend the graphene channels as shown in Fig. \ref{fig1}f. To fabricate the gold tunnel junctions (Fig. \ref{fig2}a), we used the same lithography method but exposed bow-tie shaped gold bridges. After BOE etching to suspend the gold bridges, they were cooled down to low-temperature and electromigrated \cite{Island11}.\\ \\

\textbf{Electromigration and Joule Annealing} To create mechanically-tunable gold tunnel junctions we used low-temperature electromigration to introduce nm-sized gaps in bow-tie shaped suspended gold bridges. We used a feedback-controlled ramp up of $V_{B}$ across the device while monitoring its resistance. The details were reported previously\cite{McRae17}. To remove impurities in suspended graphene channels after their cool down, we use a Joule annealing method. This method also consists in ramping up a $V_{B}$ across the device, but without any feeback loop (Fig. \ref{fig3}a). The voltage is ramped up until a desired power $P = IV_{B} $ is achieved, and then held constant for 10 minutes to anneal the device. Then, we acquired $G-V_{G}$ transconductance (Figs. \ref{fig3}b and S5d) to monitor the cleanliness of the device. We repeated this procedure with increasingly high $P\sim \mu$W until the device had the desired $G-V_{G}$ characteristics.\\ \\

\textbf{Theoretical calculations}
The complete derivation of the theoretical model is presented in Supplementary section 1, and discussed in detail in a previous work \cite{McRae19}. We mention a few key model parameters used to simulate Devices 1 and 2. In the device's $x$ - $y$ coordinates, the strain tensor $\bar{\bm{\varepsilon}}$ has elements $\varepsilon_{xx}=\varepsilon_{\text{tot}}$, $\varepsilon_{yy}=-\nu \varepsilon_{\text{tot}}$ and $\varepsilon_{xy}=\varepsilon_{yx}=0$, where $\nu=0.165$ is the Poisson ratio \cite{Naumis17}. The scalar gauge potential is then given by $\phi_{\varepsilon} = g_{\varepsilon}(1-\nu)\varepsilon_{\text{tot}}$, where\cite{Choi10} $g_{\varepsilon}\approx 3.0$~eV. The vector potentials are $\bm{A}_{i}= -\bar{\bm{\varepsilon}}\bm{K}_{i} + \bm{A}_{\text{hop}}$, where the first term comes from the movement of the FBZ corners $\bm{K}_{i}$. The second terms is due to the modification of the lattice's nearest-neighbor hopping amplitudes, $\bm{A}_{\text{hop}}=\frac{\beta\varepsilon_{\text{tot}}(1+\nu)}{2a} (\cos3\theta, \sin3\theta)$ where $\beta \approx 2.5$ \cite{Naumis17}. Although not visible in Fig. 1d, the $\bm{A}_{\text{i}}$ also displace slightly the Dirac points (band intersections) away from the corners of the FBZ (Supplementary Fig. S1).
\\ \\
\textbf{Data Availability} The data that support the findings  of this study are included in the Figures of the main text and Supplementary information. They are also available from the corresponding author upon reasonable request. \\ \\
\textbf{Acknowledgements}\\
This work was supported by NSERC (Canada), CFI (Canada), and Concordia University. We acknowledge usage of the QNI (Quebec Nano Infrastructure) cleanroom network. \\ \\
\textbf{Author Contributions}\\
A.C.M. lead the fabrication of the instrumentation, most devices, acquired most of the data presented, and contributed to all other aspects of the work. G. W. lead the data analysis and contributed to all others aspects of the work. L. H. contributed to the data analysis and modeling. S. Y. fabricated device T2 and acquired the data in Fig. \ref{fig2}e. V.T. fabricated device T1 and acquired the data in Fig. \ref{fig2}d. A.R.C designed, supervised, and made significant contributions to all aspects of the work. A.R.C. wrote the manuscript with comments and inputs from all of the authors. \\
\textbf{Competing interests} The authors declare no competing interest. \newline \newline
\textbf{Supplementary information} The online version contains supplementary material, including extensive additional data and the theoretical model derivation.
\\
\\

\bibliography{ms}

\providecommand{\latin}[1]{#1}
\makeatletter
\providecommand{\doi}
  {\begingroup\let\do\@makeother\dospecials
  \catcode`\{=1 \catcode`\}=2 \doi@aux}
\providecommand{\doi@aux}[1]{\endgroup\texttt{#1}}
\makeatother
\providecommand*\mcitethebibliography{\thebibliography}
\csname @ifundefined\endcsname{endmcitethebibliography}
  {\let\endmcitethebibliography\endthebibliography}{}
\begin{mcitethebibliography}{62}
\providecommand*\natexlab[1]{#1}
\providecommand*\mciteSetBstSublistMode[1]{}
\providecommand*\mciteSetBstMaxWidthForm[2]{}
\providecommand*\mciteBstWouldAddEndPuncttrue
  {\def\EndOfBibitem{\unskip.}}
\providecommand*\mciteBstWouldAddEndPunctfalse
  {\let\EndOfBibitem\relax}
\providecommand*\mciteSetBstMidEndSepPunct[3]{}
\providecommand*\mciteSetBstSublistLabelBeginEnd[3]{}
\providecommand*\EndOfBibitem{}
\mciteSetBstSublistMode{f}
\mciteSetBstMaxWidthForm{subitem}{(\alph{mcitesubitemcount})}
\mciteSetBstSublistLabelBeginEnd
  {\mcitemaxwidthsubitemform\space}
  {\relax}
  {\relax}

\bibitem[Alfieri \latin{et~al.}(2023)Alfieri, Anantharaman, Zhang, and
  Jariwala]{Alfieri23}
Alfieri,~A.; Anantharaman,~S.~B.; Zhang,~H.~Q.; Jariwala,~D. Nanomaterials for
  Quantum Information Science and Engineering. \emph{Adv. Mater.}
  \textbf{2023}, \emph{35}\relax
\mciteBstWouldAddEndPuncttrue
\mciteSetBstMidEndSepPunct{\mcitedefaultmidpunct}
{\mcitedefaultendpunct}{\mcitedefaultseppunct}\relax
\EndOfBibitem
\bibitem[Banszerus \latin{et~al.}(2021)Banszerus, Hecker, Icking, Trellenkamp,
  Lentz, Neumaier, Watanabe, Taniguchi, Volk, and Stampfer]{Banszerus21}
Banszerus,~L.; Hecker,~K.; Icking,~E.; Trellenkamp,~S.; Lentz,~F.;
  Neumaier,~D.; Watanabe,~K.; Taniguchi,~T.; Volk,~C.; Stampfer,~C. Pulsed-gate
  spectroscopy of single-electron spin states in bilayer graphene quantum dots.
  \emph{Phys. Rev. B} \textbf{2021}, \emph{103}, L081404\relax
\mciteBstWouldAddEndPuncttrue
\mciteSetBstMidEndSepPunct{\mcitedefaultmidpunct}
{\mcitedefaultendpunct}{\mcitedefaultseppunct}\relax
\EndOfBibitem
\bibitem[Liu and Hersam(2019)Liu, and Hersam]{Liu19}
Liu,~X.~L.; Hersam,~M.~C. 2D materials for quantum information science.
  \emph{Nat. Rev. Mater.} \textbf{2019}, \emph{4}, 669--684\relax
\mciteBstWouldAddEndPuncttrue
\mciteSetBstMidEndSepPunct{\mcitedefaultmidpunct}
{\mcitedefaultendpunct}{\mcitedefaultseppunct}\relax
\EndOfBibitem
\bibitem[Pal \latin{et~al.}(2023)Pal, Zhang, Chavan, Agashiwala, Yeh, Cao, and
  Banerjee]{Pal23}
Pal,~A.; Zhang,~S.; Chavan,~T.; Agashiwala,~K.; Yeh,~C.~H.; Cao,~W.;
  Banerjee,~K. Quantum-Engineered Devices Based on 2D Materials for
  Next-Generation Information Processing and Storage. \emph{Adv. Mater.}
  \textbf{2023}, \emph{35}\relax
\mciteBstWouldAddEndPuncttrue
\mciteSetBstMidEndSepPunct{\mcitedefaultmidpunct}
{\mcitedefaultendpunct}{\mcitedefaultseppunct}\relax
\EndOfBibitem
\bibitem[Wu \latin{et~al.}(2018)Wu, Fatemi, Gibson, Watanabe, Taniguchi, Cava,
  and Jarillo-Herrero]{Wu18}
Wu,~S.~F.; Fatemi,~V.; Gibson,~Q.~D.; Watanabe,~K.; Taniguchi,~T.; Cava,~R.~J.;
  Jarillo-Herrero,~P. Observation of the quantum spin Hall effect up to 100
  kelvin in a monolayer crystal. \emph{Science} \textbf{2018}, \emph{359},
  76--79\relax
\mciteBstWouldAddEndPuncttrue
\mciteSetBstMidEndSepPunct{\mcitedefaultmidpunct}
{\mcitedefaultendpunct}{\mcitedefaultseppunct}\relax
\EndOfBibitem
\bibitem[Hanakata \latin{et~al.}(2018)Hanakata, Rodin, Park, Campbell, and
  Neto]{Hanakata18}
Hanakata,~P.~Z.; Rodin,~A.~S.; Park,~H.~S.; Campbell,~D.~K.; Neto,~A. H.~C.
  Strain-induced gauge and Rashba fields in ferroelectric Rashba lead
  chalcogenide PbX monolayers (X = S, Se, Te). \emph{Phys. Rev. B}
  \textbf{2018}, \emph{97}\relax
\mciteBstWouldAddEndPuncttrue
\mciteSetBstMidEndSepPunct{\mcitedefaultmidpunct}
{\mcitedefaultendpunct}{\mcitedefaultseppunct}\relax
\EndOfBibitem
\bibitem[Molle \latin{et~al.}(2017)Molle, Goldberger, Houssa, Xu, Zhang, and
  Akinwande]{Molle17}
Molle,~A.; Goldberger,~J.; Houssa,~M.; Xu,~Y.; Zhang,~S.~C.; Akinwande,~D.
  Buckled two-dimensional Xene sheets. \emph{Nat. Mater.} \textbf{2017},
  \emph{16}, 163--169\relax
\mciteBstWouldAddEndPuncttrue
\mciteSetBstMidEndSepPunct{\mcitedefaultmidpunct}
{\mcitedefaultendpunct}{\mcitedefaultseppunct}\relax
\EndOfBibitem
\bibitem[Li \latin{et~al.}(2020)Li, Su, Ren, and He]{Li20}
Li,~S.~Y.; Su,~Y.; Ren,~Y.~N.; He,~L. Valley Polarization and Inversion in
  Strained Graphene via Pseudo-Landau Levels, Valley Splitting of Real Landau
  Levels, and Confined States. \emph{Phys. Rev. Lett.} \textbf{2020},
  \emph{124}, 106802\relax
\mciteBstWouldAddEndPuncttrue
\mciteSetBstMidEndSepPunct{\mcitedefaultmidpunct}
{\mcitedefaultendpunct}{\mcitedefaultseppunct}\relax
\EndOfBibitem
\bibitem[Settnes \latin{et~al.}(2017)Settnes, Garcia, and Roche]{Settnes17}
Settnes,~M.; Garcia,~J.~H.; Roche,~S. Valley-polarized quantum transport
  generated by gauge fields in graphene. \emph{2D Mater.} \textbf{2017},
  \emph{4}, 031006\relax
\mciteBstWouldAddEndPuncttrue
\mciteSetBstMidEndSepPunct{\mcitedefaultmidpunct}
{\mcitedefaultendpunct}{\mcitedefaultseppunct}\relax
\EndOfBibitem
\bibitem[Schaibley \latin{et~al.}(2016)Schaibley, Yu, Clark, Rivera, Ross,
  Seyler, Yao, and Xu]{Schaibley16}
Schaibley,~J.~R.; Yu,~H.~Y.; Clark,~G.; Rivera,~P.; Ross,~J.~S.; Seyler,~K.~L.;
  Yao,~W.; Xu,~X.~D. Valleytronics in 2D materials. \emph{Nat. Rev. Mater.}
  \textbf{2016}, \emph{1}, 16055\relax
\mciteBstWouldAddEndPuncttrue
\mciteSetBstMidEndSepPunct{\mcitedefaultmidpunct}
{\mcitedefaultendpunct}{\mcitedefaultseppunct}\relax
\EndOfBibitem
\bibitem[Guinea \latin{et~al.}(2010)Guinea, Katsnelson, and Geim]{Guinea10}
Guinea,~F.; Katsnelson,~M.~I.; Geim,~A.~K. Energy gaps and a zero-field quantum
  Hall effect in graphene by strain engineering. \emph{Nat. Phys.}
  \textbf{2010}, \emph{6}, 30\relax
\mciteBstWouldAddEndPuncttrue
\mciteSetBstMidEndSepPunct{\mcitedefaultmidpunct}
{\mcitedefaultendpunct}{\mcitedefaultseppunct}\relax
\EndOfBibitem
\bibitem[Kapfer \latin{et~al.}(2023)Kapfer, Jessen, Eisele, Fu, Danielsen,
  Darlington, Moore, Finney, Marchese, Hsieh, Majchrzak, Jiang, Biswas, Dudin,
  Avila, Watanabe, Taniguchi, Ulstrup, Boggild, Schuck, Basov, Hone, and
  Dean]{Kapfer23}
Kapfer,~M. \latin{et~al.}  Programming twist angle and strain profiles in 2D
  materials. \emph{Science} \textbf{2023}, \emph{381}, 677--681\relax
\mciteBstWouldAddEndPuncttrue
\mciteSetBstMidEndSepPunct{\mcitedefaultmidpunct}
{\mcitedefaultendpunct}{\mcitedefaultseppunct}\relax
\EndOfBibitem
\bibitem[Kim \latin{et~al.}(2023)Kim, Haque, Hsieh, Nahid, Zarin, Jeong, So,
  Park, and Nam]{Kim23}
Kim,~J.~M.; Haque,~M.~F.; Hsieh,~E.~Y.; Nahid,~S.~M.; Zarin,~I.; Jeong,~K.~Y.;
  So,~J.~P.; Park,~H.~G.; Nam,~S. Strain Engineering of Low-Dimensional
  Materials for Emerging Quantum Phenomena and Functionalities. \emph{Adv.
  Mater.} \textbf{2023}, \emph{35}, 2107362\relax
\mciteBstWouldAddEndPuncttrue
\mciteSetBstMidEndSepPunct{\mcitedefaultmidpunct}
{\mcitedefaultendpunct}{\mcitedefaultseppunct}\relax
\EndOfBibitem
\bibitem[Rakib \latin{et~al.}(2022)Rakib, Pochet, Ertekin, and
  Johnson]{Rakib22}
Rakib,~T.; Pochet,~P.; Ertekin,~E.; Johnson,~H.~T. Corrugation-driven symmetry
  breaking in magic-angle twisted bilayer graphene. \emph{Communications
  Physics} \textbf{2022}, \emph{5}, 242\relax
\mciteBstWouldAddEndPuncttrue
\mciteSetBstMidEndSepPunct{\mcitedefaultmidpunct}
{\mcitedefaultendpunct}{\mcitedefaultseppunct}\relax
\EndOfBibitem
\bibitem[Khanjani and Moghaddam(2018)Khanjani, and Moghaddam]{Khanjani18}
Khanjani,~H.; Moghaddam,~A.~G. Anomalous quantum interference effects in
  graphene SNS junctions due to strain-induced gauge fields. \emph{Phys. Rev.
  B} \textbf{2018}, \emph{98}, 195421\relax
\mciteBstWouldAddEndPuncttrue
\mciteSetBstMidEndSepPunct{\mcitedefaultmidpunct}
{\mcitedefaultendpunct}{\mcitedefaultseppunct}\relax
\EndOfBibitem
\bibitem[Zhang \latin{et~al.}(2023)Zhang, He, Liu, Dai, Liu, Chen, Wu, Zhu, and
  Yang]{Zhang23}
Zhang,~X.~M.; He,~T.~L.; Liu,~Y.; Dai,~X.~F.; Liu,~G.~D.; Chen,~C.; Wu,~W.~K.;
  Zhu,~J.~J.; Yang,~S. Y.~A. Magnetic Real Chern Insulator in 2D Metal-Organic
  Frameworks. \emph{Nano Lett.} \textbf{2023}, 7358--7363\relax
\mciteBstWouldAddEndPuncttrue
\mciteSetBstMidEndSepPunct{\mcitedefaultmidpunct}
{\mcitedefaultendpunct}{\mcitedefaultseppunct}\relax
\EndOfBibitem
\bibitem[Pantaleón \latin{et~al.}(2021)Pantaleón, Low, and
  Guinea]{Pantaleon21}
Pantaleón,~P.~A.; Low,~T.; Guinea,~F. Tunable large Berry dipole in strained
  twisted bilayer graphene. \emph{Phys. Rev. B} \textbf{2021}, \emph{103},
  205403\relax
\mciteBstWouldAddEndPuncttrue
\mciteSetBstMidEndSepPunct{\mcitedefaultmidpunct}
{\mcitedefaultendpunct}{\mcitedefaultseppunct}\relax
\EndOfBibitem
\bibitem[Du \latin{et~al.}(2021)Du, Hasan, Castellanos-Gomez, Liu, Yao, Lau,
  and Sun]{Du21}
Du,~L.~J.; Hasan,~T.; Castellanos-Gomez,~A.; Liu,~G.~B.; Yao,~Y.~G.;
  Lau,~C.~N.; Sun,~Z.~P. Engineering symmetry breaking in 2D layered materials.
  \emph{Nat. Rev. Phys.} \textbf{2021}, \emph{3}, 193--206\relax
\mciteBstWouldAddEndPuncttrue
\mciteSetBstMidEndSepPunct{\mcitedefaultmidpunct}
{\mcitedefaultendpunct}{\mcitedefaultseppunct}\relax
\EndOfBibitem
\bibitem[Moulsdale \latin{et~al.}(2020)Moulsdale, Knothe, and
  Fal'ko]{Moulsdale20}
Moulsdale,~C.; Knothe,~A.; Fal'ko,~V. Engineering of the topological magnetic
  moment of electrons in bilayer graphene using strain and electrical bias.
  \emph{Phys. Rev. B} \textbf{2020}, \emph{101}, 085118\relax
\mciteBstWouldAddEndPuncttrue
\mciteSetBstMidEndSepPunct{\mcitedefaultmidpunct}
{\mcitedefaultendpunct}{\mcitedefaultseppunct}\relax
\EndOfBibitem
\bibitem[Mutch \latin{et~al.}(2019)Mutch, Chen, Went, Qian, Wilson, Andreev,
  Chen, and Chu]{Mutch19}
Mutch,~J.; Chen,~W.~C.; Went,~P.; Qian,~T.~M.; Wilson,~I.~Z.; Andreev,~A.;
  Chen,~C.~C.; Chu,~J.~H. Evidence for a strain-tuned topological phase
  transition in ZrTe5. \emph{Science Advances} \textbf{2019}, \emph{5}\relax
\mciteBstWouldAddEndPuncttrue
\mciteSetBstMidEndSepPunct{\mcitedefaultmidpunct}
{\mcitedefaultendpunct}{\mcitedefaultseppunct}\relax
\EndOfBibitem
\bibitem[Cenker \latin{et~al.}(2022)Cenker, Sivakumar, Xie, Miller, Thijssen,
  Liu, Dismukes, Fonseca, Anderson, Zhu, Roy, Xiao, Chu, Cao, and Xu]{Cenker22}
Cenker,~J.; Sivakumar,~S.; Xie,~K.~C.; Miller,~A.; Thijssen,~P.; Liu,~Z.~Y.;
  Dismukes,~A.; Fonseca,~J.; Anderson,~E.; Zhu,~X.~Y.; Roy,~X.; Xiao,~D.;
  Chu,~J.~H.; Cao,~T.; Xu,~X.~D. Reversible strain-induced magnetic phase
  transition in a van der Waals magnet. \emph{Nat. Nanotechnol.} \textbf{2022},
  \emph{17}, 256\relax
\mciteBstWouldAddEndPuncttrue
\mciteSetBstMidEndSepPunct{\mcitedefaultmidpunct}
{\mcitedefaultendpunct}{\mcitedefaultseppunct}\relax
\EndOfBibitem
\bibitem[Li \latin{et~al.}(2019)Li, Jiang, Sivadas, Wang, Xu, Weber,
  Goldberger, Watanabe, Taniguchi, Fennie, Mak, and Shan]{Li19}
Li,~T.~X.; Jiang,~S.~W.; Sivadas,~N.; Wang,~Z.~F.; Xu,~Y.; Weber,~D.;
  Goldberger,~J.~E.; Watanabe,~K.; Taniguchi,~T.; Fennie,~C.~J.; Mak,~K.~F.;
  Shan,~J. Pressure-controlled interlayer magnetism in atomically thin CrI.
  \emph{Nat. Mater.} \textbf{2019}, \emph{18}, 1303--1308\relax
\mciteBstWouldAddEndPuncttrue
\mciteSetBstMidEndSepPunct{\mcitedefaultmidpunct}
{\mcitedefaultendpunct}{\mcitedefaultseppunct}\relax
\EndOfBibitem
\bibitem[Burch \latin{et~al.}(2018)Burch, Mandrus, and Park]{Burch18}
Burch,~K.~S.; Mandrus,~D.; Park,~J.~G. Magnetism in two-dimensional van der
  Waals materials. \emph{Nature} \textbf{2018}, \emph{563}, 47--52\relax
\mciteBstWouldAddEndPuncttrue
\mciteSetBstMidEndSepPunct{\mcitedefaultmidpunct}
{\mcitedefaultendpunct}{\mcitedefaultseppunct}\relax
\EndOfBibitem
\bibitem[Fogler \latin{et~al.}(2008)Fogler, Guinea, and Katsnelson]{Fogler08}
Fogler,~M.~M.; Guinea,~F.; Katsnelson,~M.~I. Pseudomagnetic fields and
  ballistic transport in a suspended graphene sheet. \emph{Phys. Rev. Lett.}
  \textbf{2008}, \emph{101}, 226804\relax
\mciteBstWouldAddEndPuncttrue
\mciteSetBstMidEndSepPunct{\mcitedefaultmidpunct}
{\mcitedefaultendpunct}{\mcitedefaultseppunct}\relax
\EndOfBibitem
\bibitem[McRae \latin{et~al.}(2019)McRae, Wei, and Champagne]{McRae19}
McRae,~A.~C.; Wei,~G.; Champagne,~A.~R. Graphene Quantum Strain Transistors.
  \emph{Phys. Rev. Appl.} \textbf{2019}, \emph{11}, 054019\relax
\mciteBstWouldAddEndPuncttrue
\mciteSetBstMidEndSepPunct{\mcitedefaultmidpunct}
{\mcitedefaultendpunct}{\mcitedefaultseppunct}\relax
\EndOfBibitem
\bibitem[Kazmierczak \latin{et~al.}(2021)Kazmierczak, Van~Winkle, Ophus,
  Bustillo, Carr, Brown, Ciston, Taniguchi, Watanabe, and
  Bediako]{Kazmierczak21}
Kazmierczak,~N.~P.; Van~Winkle,~M.; Ophus,~C.; Bustillo,~K.~C.; Carr,~S.;
  Brown,~H.~G.; Ciston,~J.; Taniguchi,~T.; Watanabe,~K.; Bediako,~D.~K. Strain
  fields in twisted bilayer graphene. \emph{Nat. Mater.} \textbf{2021},
  \emph{20}, 956\relax
\mciteBstWouldAddEndPuncttrue
\mciteSetBstMidEndSepPunct{\mcitedefaultmidpunct}
{\mcitedefaultendpunct}{\mcitedefaultseppunct}\relax
\EndOfBibitem
\bibitem[Dirnberger \latin{et~al.}(2021)Dirnberger, Ziegler, Faria, Bushati,
  Taniguchi, Watanabe, Fabian, Bougeard, Chernikov, and Menon]{Dirnberger21}
Dirnberger,~F.; Ziegler,~J.~D.; Faria,~P.~E.; Bushati,~R.; Taniguchi,~T.;
  Watanabe,~K.; Fabian,~J.; Bougeard,~D.; Chernikov,~A.; Menon,~V.~M. Quasi-1D
  exciton channels in strain-engineered 2D materials. \emph{Sci. Adv.}
  \textbf{2021}, \emph{7}, eabj3066\relax
\mciteBstWouldAddEndPuncttrue
\mciteSetBstMidEndSepPunct{\mcitedefaultmidpunct}
{\mcitedefaultendpunct}{\mcitedefaultseppunct}\relax
\EndOfBibitem
\bibitem[Mesple \latin{et~al.}(2021)Mesple, Missaoui, Cea, Huder, Guinea,
  Trambly~de Laissardière, Chapelier, and Renard]{Mesple21}
Mesple,~F.; Missaoui,~A.; Cea,~T.; Huder,~L.; Guinea,~F.; Trambly~de
  Laissardière,~G.; Chapelier,~C.; Renard,~V.~T. Heterostrain Determines Flat
  Bands in Magic-Angle Twisted Graphene Layers. \emph{Phys. Rev. Lett.}
  \textbf{2021}, \emph{127}, 126405\relax
\mciteBstWouldAddEndPuncttrue
\mciteSetBstMidEndSepPunct{\mcitedefaultmidpunct}
{\mcitedefaultendpunct}{\mcitedefaultseppunct}\relax
\EndOfBibitem
\bibitem[Shi \latin{et~al.}(2020)Shi, Zhan, Qi, Huang, van Veen, Silva-Guillen,
  Zhang, Li, Xie, Ji, Katsnelson, Yuan, Qin, and Zhang]{Shi20}
Shi,~H.~H.; Zhan,~Z.; Qi,~Z.~K.; Huang,~K.~X.; van Veen,~E.;
  Silva-Guillen,~J.~A.; Zhang,~R.~X.; Li,~P.~J.; Xie,~K.; Ji,~H.~X.;
  Katsnelson,~M.~I.; Yuan,~S.~J.; Qin,~S.~Y.; Zhang,~Z.~Y. Large-area,
  periodic, and tunable intrinsic pseudo-magnetic fields in low-angle twisted
  bilayer graphene. \emph{Nat. Commun.} \textbf{2020}, \emph{11}, 371\relax
\mciteBstWouldAddEndPuncttrue
\mciteSetBstMidEndSepPunct{\mcitedefaultmidpunct}
{\mcitedefaultendpunct}{\mcitedefaultseppunct}\relax
\EndOfBibitem
\bibitem[Banerjee \latin{et~al.}(2020)Banerjee, Nguyen, Granzier-Nakajima,
  Pabbi, Lherbier, Binion, Charlier, Terrones, and Hudson]{Banerjee20}
Banerjee,~R.; Nguyen,~V.~H.; Granzier-Nakajima,~T.; Pabbi,~L.; Lherbier,~A.;
  Binion,~A.~R.; Charlier,~J.~C.; Terrones,~M.; Hudson,~E.~W. Strain Modulated
  Superlattices in Graphene. \emph{Nano Lett.} \textbf{2020}, \emph{20},
  3113--3121\relax
\mciteBstWouldAddEndPuncttrue
\mciteSetBstMidEndSepPunct{\mcitedefaultmidpunct}
{\mcitedefaultendpunct}{\mcitedefaultseppunct}\relax
\EndOfBibitem
\bibitem[Nigge \latin{et~al.}(2019)Nigge, Qu, Lantagne-Hurtubise, Marsell,
  Link, Tom, Zonno, Michiardi, Schneider, Zhdanovich, Levy, Starke, Gutierrez,
  Bonn, Burke, Franz, and Damascelli]{Nigge19}
Nigge,~P. \latin{et~al.}  Room temperature strain-induced Landau levels in
  graphene on a wafer-scale platform. \emph{Sci. Adv.} \textbf{2019}, \emph{5},
  eaaw5593\relax
\mciteBstWouldAddEndPuncttrue
\mciteSetBstMidEndSepPunct{\mcitedefaultmidpunct}
{\mcitedefaultendpunct}{\mcitedefaultseppunct}\relax
\EndOfBibitem
\bibitem[Liu \latin{et~al.}(2018)Liu, Rodrigues, Luo, Li, Carvalho, Yang,
  Laksono, Lu, Bao, Xu, Tan, Qiu, Sow, Feng, Neto, Adam, Lu, and Loh]{Liu18}
Liu,~Y.~P. \latin{et~al.}  Tailoring sample-wide pseudo-magnetic fields on a
  graphene-black phosphorus heterostructure. \emph{Nat. Nanotechnol.}
  \textbf{2018}, \emph{13}, 828\relax
\mciteBstWouldAddEndPuncttrue
\mciteSetBstMidEndSepPunct{\mcitedefaultmidpunct}
{\mcitedefaultendpunct}{\mcitedefaultseppunct}\relax
\EndOfBibitem
\bibitem[Goldsche \latin{et~al.}(2018)Goldsche, Sonntag, Khodkov, Verbiest,
  Reichardt, Neumann, Ouaj, von~den Driesch, Buca, and Stampfer]{Goldsche18}
Goldsche,~M.; Sonntag,~J.; Khodkov,~T.; Verbiest,~G.~J.; Reichardt,~S.;
  Neumann,~C.; Ouaj,~T.; von~den Driesch,~N.; Buca,~D.; Stampfer,~C. Tailoring
  Mechanically Tunable Strain Fields in Graphene. \emph{Nano Lett.}
  \textbf{2018}, \emph{18}, 1707\relax
\mciteBstWouldAddEndPuncttrue
\mciteSetBstMidEndSepPunct{\mcitedefaultmidpunct}
{\mcitedefaultendpunct}{\mcitedefaultseppunct}\relax
\EndOfBibitem
\bibitem[Jiang \latin{et~al.}(2017)Jiang, Mao, Duan, Lai, Watanabe, Taniguchi,
  and Andrei]{Jiang17}
Jiang,~Y.; Mao,~J.; Duan,~J.; Lai,~X.; Watanabe,~K.; Taniguchi,~T.;
  Andrei,~E.~Y. Visualizing Strain-Induced Pseudomagnetic Fields in Graphene
  through an hBN Magnifying Glass. \emph{Nano Lett.} \textbf{2017}, \emph{17},
  2839\relax
\mciteBstWouldAddEndPuncttrue
\mciteSetBstMidEndSepPunct{\mcitedefaultmidpunct}
{\mcitedefaultendpunct}{\mcitedefaultseppunct}\relax
\EndOfBibitem
\bibitem[Levy \latin{et~al.}(2010)Levy, Burke, Meaker, Panlasigui, Zettl,
  Guinea, Castro~Neto, and Crommie]{Levy10}
Levy,~N.; Burke,~S.~A.; Meaker,~K.~L.; Panlasigui,~M.; Zettl,~A.; Guinea,~F.;
  Castro~Neto,~A.~H.; Crommie,~M.~F. Strain-Induced Pseudo–Magnetic Fields
  Greater Than 300 Tesla in Graphene Nanobubbles. \emph{Science} \textbf{2010},
  \emph{329}, 544\relax
\mciteBstWouldAddEndPuncttrue
\mciteSetBstMidEndSepPunct{\mcitedefaultmidpunct}
{\mcitedefaultendpunct}{\mcitedefaultseppunct}\relax
\EndOfBibitem
\bibitem[Wang \latin{et~al.}(2021)Wang, Baumgartner, Makk, Zihlmann, Varghese,
  Indolese, Watanabe, Taniguchi, and Schonenberger]{Wang21}
Wang,~L.; Baumgartner,~A.; Makk,~P.; Zihlmann,~S.; Varghese,~B.~S.;
  Indolese,~D.~I.; Watanabe,~K.; Taniguchi,~T.; Schonenberger,~C. Global
  strain-induced scalar potential in graphene devices. \emph{Commun. Phys.}
  \textbf{2021}, \emph{4}, 147\relax
\mciteBstWouldAddEndPuncttrue
\mciteSetBstMidEndSepPunct{\mcitedefaultmidpunct}
{\mcitedefaultendpunct}{\mcitedefaultseppunct}\relax
\EndOfBibitem
\bibitem[Wang \latin{et~al.}(2020)Wang, Wang, Liang, Ma, Xu, Liu, Zhang,
  Admasu, Cheong, Wang, Chen, Liu, Cheng, Ji, and Miao]{Wang20_2}
Wang,~Y.; Wang,~C.; Liang,~S.~J.; Ma,~Z.~C.; Xu,~K.; Liu,~X.~W.; Zhang,~L.~L.;
  Admasu,~A.~S.; Cheong,~S.~W.; Wang,~L.~Z.; Chen,~M.~Y.; Liu,~Z.~L.;
  Cheng,~B.; Ji,~W.; Miao,~F. Strain-Sensitive Magnetization Reversal of a van
  der Waals Magnet. \emph{Adv. Mater.} \textbf{2020}, \emph{32}, 2004533\relax
\mciteBstWouldAddEndPuncttrue
\mciteSetBstMidEndSepPunct{\mcitedefaultmidpunct}
{\mcitedefaultendpunct}{\mcitedefaultseppunct}\relax
\EndOfBibitem
\bibitem[Zhang \latin{et~al.}(2019)Zhang, Kim, Gilbert, and Mason]{Zhang19}
Zhang,~Y.~J.; Kim,~Y.; Gilbert,~M.~J.; Mason,~N. Magnetotransport in a strain
  superlattice of graphene. \emph{Appl. Phys. Lett.} \textbf{2019},
  \emph{115}\relax
\mciteBstWouldAddEndPuncttrue
\mciteSetBstMidEndSepPunct{\mcitedefaultmidpunct}
{\mcitedefaultendpunct}{\mcitedefaultseppunct}\relax
\EndOfBibitem
\bibitem[De~Sanctis \latin{et~al.}(2018)De~Sanctis, Mehew, Alkhalifa, Withers,
  Craciun, and Russo]{De_Sanctis18}
De~Sanctis,~A.; Mehew,~J.~D.; Alkhalifa,~S.; Withers,~F.; Craciun,~M.~F.;
  Russo,~S. Strain-Engineering of Twist-Angle in Graphene/hBN Superlattice
  Devices. \emph{Nano Lett.} \textbf{2018}, \emph{18}, 7919--7926\relax
\mciteBstWouldAddEndPuncttrue
\mciteSetBstMidEndSepPunct{\mcitedefaultmidpunct}
{\mcitedefaultendpunct}{\mcitedefaultseppunct}\relax
\EndOfBibitem
\bibitem[Guan and Du(2017)Guan, and Du]{Guan17}
Guan,~F.; Du,~X. Random Gauge Field Scattering in Monolayer Graphene.
  \emph{Nano Lett.} \textbf{2017}, \emph{17}, 7009--7014\relax
\mciteBstWouldAddEndPuncttrue
\mciteSetBstMidEndSepPunct{\mcitedefaultmidpunct}
{\mcitedefaultendpunct}{\mcitedefaultseppunct}\relax
\EndOfBibitem
\bibitem[Shioya \latin{et~al.}(2015)Shioya, Russo, Yamamoto, Craciun, and
  Tarucha]{Shioya15}
Shioya,~H.; Russo,~S.; Yamamoto,~M.; Craciun,~M.~F.; Tarucha,~S. Electron
  States of Uniaxially Strained Graphene. \emph{Nano Lett.} \textbf{2015},
  \emph{15}, 7943--7948\relax
\mciteBstWouldAddEndPuncttrue
\mciteSetBstMidEndSepPunct{\mcitedefaultmidpunct}
{\mcitedefaultendpunct}{\mcitedefaultseppunct}\relax
\EndOfBibitem
\bibitem[Dean \latin{et~al.}(2010)Dean, Young, Meric, Lee, Wang, Sorgenfrei,
  Watanabe, Taniguchi, Kim, Shepard, and Hone]{Dean10}
Dean,~C.~R.; Young,~A.~F.; Meric,~I.; Lee,~C.; Wang,~L.; Sorgenfrei,~S.;
  Watanabe,~K.; Taniguchi,~T.; Kim,~P.; Shepard,~K.~L.; Hone,~J. Boron nitride
  substrates for high-quality graphene electronics. \emph{Nat. Nanotechnol.}
  \textbf{2010}, \emph{5}, 722--726\relax
\mciteBstWouldAddEndPuncttrue
\mciteSetBstMidEndSepPunct{\mcitedefaultmidpunct}
{\mcitedefaultendpunct}{\mcitedefaultseppunct}\relax
\EndOfBibitem
\bibitem[Miao \latin{et~al.}(2021)Miao, Liang, and Cheng]{Miao21}
Miao,~F.; Liang,~S.~J.; Cheng,~B. Straintronics with van der Waals materials.
  \emph{Npj Quantum Mater.} \textbf{2021}, \emph{6}, 59\relax
\mciteBstWouldAddEndPuncttrue
\mciteSetBstMidEndSepPunct{\mcitedefaultmidpunct}
{\mcitedefaultendpunct}{\mcitedefaultseppunct}\relax
\EndOfBibitem
\bibitem[Naumis \latin{et~al.}(2017)Naumis, Barraza-Lopez, Oliva-Leyva, and
  Terrones]{Naumis17}
Naumis,~G.~G.; Barraza-Lopez,~S.; Oliva-Leyva,~M.; Terrones,~H. Electronic and
  optical properties of strained graphene and other strained 2D materials: a
  review. \emph{Rep. Prog. Phys.} \textbf{2017}, \emph{80}, 096501\relax
\mciteBstWouldAddEndPuncttrue
\mciteSetBstMidEndSepPunct{\mcitedefaultmidpunct}
{\mcitedefaultendpunct}{\mcitedefaultseppunct}\relax
\EndOfBibitem
\bibitem[van Oudenaarden \latin{et~al.}(1998)van Oudenaarden, Devoret, Nazarov,
  and Mooij]{van_Oudenaarden98}
van Oudenaarden,~A.; Devoret,~M.~H.; Nazarov,~Y.~V.; Mooij,~J.~E.
  Magneto-electric Aharonov-Bohm effect in metal rings. \emph{Nature}
  \textbf{1998}, \emph{391}, 768--770\relax
\mciteBstWouldAddEndPuncttrue
\mciteSetBstMidEndSepPunct{\mcitedefaultmidpunct}
{\mcitedefaultendpunct}{\mcitedefaultseppunct}\relax
\EndOfBibitem
\bibitem[McRae \latin{et~al.}(2017)McRae, Tayari, Porter, and
  Champagne]{McRae17}
McRae,~A.~C.; Tayari,~V.; Porter,~J.~M.; Champagne,~A.~R. Giant electron-hole
  transport asymmetry in ultra-short quantum transistors. \emph{Nat. Commun.}
  \textbf{2017}, \emph{8}, 15491\relax
\mciteBstWouldAddEndPuncttrue
\mciteSetBstMidEndSepPunct{\mcitedefaultmidpunct}
{\mcitedefaultendpunct}{\mcitedefaultseppunct}\relax
\EndOfBibitem
\bibitem[Chaves \latin{et~al.}(2014)Chaves, Jimenez, Cummings, and
  Roche]{Chaves14}
Chaves,~F.~A.; Jimenez,~D.; Cummings,~A.~W.; Roche,~S. Physical model of the
  contact resistivity of metal-graphene junctions. \emph{J. Appl. Phys.}
  \textbf{2014}, \emph{115}, 164513\relax
\mciteBstWouldAddEndPuncttrue
\mciteSetBstMidEndSepPunct{\mcitedefaultmidpunct}
{\mcitedefaultendpunct}{\mcitedefaultseppunct}\relax
\EndOfBibitem
\bibitem[Heinze \latin{et~al.}(2002)Heinze, Tersoff, Martel, Derycke,
  Appenzeller, and Avouris]{Heinze02}
Heinze,~S.; Tersoff,~J.; Martel,~R.; Derycke,~V.; Appenzeller,~J.; Avouris,~P.
  Carbon nanotubes as Schottky barrier transistors. \emph{Phys. Rev. Lett.}
  \textbf{2002}, \emph{89}, 106801\relax
\mciteBstWouldAddEndPuncttrue
\mciteSetBstMidEndSepPunct{\mcitedefaultmidpunct}
{\mcitedefaultendpunct}{\mcitedefaultseppunct}\relax
\EndOfBibitem
\bibitem[Tworzydlo \latin{et~al.}(2006)Tworzydlo, Trauzettel, Titov, Rycerz,
  and Beenakker]{Tworzydlo06}
Tworzydlo,~J.; Trauzettel,~B.; Titov,~M.; Rycerz,~A.; Beenakker,~C.~W.
  Sub-Poissonian shot noise in graphene. \emph{Phys. Rev. Lett.} \textbf{2006},
  \emph{96}, 246802\relax
\mciteBstWouldAddEndPuncttrue
\mciteSetBstMidEndSepPunct{\mcitedefaultmidpunct}
{\mcitedefaultendpunct}{\mcitedefaultseppunct}\relax
\EndOfBibitem
\bibitem[Champagne \latin{et~al.}(2005)Champagne, Pasupathy, and
  Ralph]{Champagne05}
Champagne,~A.~R.; Pasupathy,~A.~N.; Ralph,~D.~C. Mechanically adjustable and
  electrically gated single-molecule transistors. \emph{Nano Lett.}
  \textbf{2005}, \emph{5}, 305--308\relax
\mciteBstWouldAddEndPuncttrue
\mciteSetBstMidEndSepPunct{\mcitedefaultmidpunct}
{\mcitedefaultendpunct}{\mcitedefaultseppunct}\relax
\EndOfBibitem
\bibitem[Choi \latin{et~al.}(2010)Choi, Jhi, and Son]{Choi10}
Choi,~S.-M.; Jhi,~S.-H.; Son,~Y.-W. Effects of strain on electronic properties
  of graphene. \emph{Phys. Rev. B} \textbf{2010}, \emph{81}, 081407\relax
\mciteBstWouldAddEndPuncttrue
\mciteSetBstMidEndSepPunct{\mcitedefaultmidpunct}
{\mcitedefaultendpunct}{\mcitedefaultseppunct}\relax
\EndOfBibitem
\bibitem[Grassano \latin{et~al.}(2020)Grassano, D'Alessandro, Pulci, Sharapov,
  Gusynin, and Varlamov]{Grassano20}
Grassano,~D.; D'Alessandro,~M.; Pulci,~O.; Sharapov,~S.~G.; Gusynin,~V.~P.;
  Varlamov,~A.~A. Work function, deformation potential, and collapse of Landau
  levels in strained graphene and silicene. \emph{Phys. Rev. B} \textbf{2020},
  \emph{101}, 245115\relax
\mciteBstWouldAddEndPuncttrue
\mciteSetBstMidEndSepPunct{\mcitedefaultmidpunct}
{\mcitedefaultendpunct}{\mcitedefaultseppunct}\relax
\EndOfBibitem
\bibitem[Yigen and Champagne(2014)Yigen, and Champagne]{Yigen14}
Yigen,~S.; Champagne,~A.~R. Wiedemann-Franz Relation and Thermal-Transistor
  Effect in Suspended Graphene. \emph{Nano Lett.} \textbf{2014}, \emph{14},
  289\relax
\mciteBstWouldAddEndPuncttrue
\mciteSetBstMidEndSepPunct{\mcitedefaultmidpunct}
{\mcitedefaultendpunct}{\mcitedefaultseppunct}\relax
\EndOfBibitem
\bibitem[Bolotin \latin{et~al.}(2008)Bolotin, Sikes, Jiang, Klima, Fudenberg,
  Hone, Kim, and Stormer]{Bolotin08}
Bolotin,~K.~I.; Sikes,~K.~J.; Jiang,~Z.; Klima,~M.; Fudenberg,~G.; Hone,~J.;
  Kim,~P.; Stormer,~H.~L. Ultrahigh electron mobility in suspended graphene.
  \emph{Solid State Commun.} \textbf{2008}, \emph{146}, 351--355\relax
\mciteBstWouldAddEndPuncttrue
\mciteSetBstMidEndSepPunct{\mcitedefaultmidpunct}
{\mcitedefaultendpunct}{\mcitedefaultseppunct}\relax
\EndOfBibitem
\bibitem[Low and Appenzeller(2009)Low, and Appenzeller]{Low09}
Low,~T.; Appenzeller,~J. Electronic transport properties of a tilted graphene
  p-n junction. \emph{Phys. Rev. B} \textbf{2009}, \emph{80}, 155406\relax
\mciteBstWouldAddEndPuncttrue
\mciteSetBstMidEndSepPunct{\mcitedefaultmidpunct}
{\mcitedefaultendpunct}{\mcitedefaultseppunct}\relax
\EndOfBibitem
\bibitem[Laitinen \latin{et~al.}(2016)Laitinen, Paraoanu, Oksanen, Craciun,
  Russo, Sonin, and Hakonen]{Laitinen16}
Laitinen,~A.; Paraoanu,~G.~S.; Oksanen,~M.; Craciun,~M.~F.; Russo,~S.;
  Sonin,~E.; Hakonen,~P. Contact doping, Klein tunneling, and asymmetry of shot
  noise in suspended graphene. \emph{Phys. Rev. B} \textbf{2016}, \emph{93},
  115413\relax
\mciteBstWouldAddEndPuncttrue
\mciteSetBstMidEndSepPunct{\mcitedefaultmidpunct}
{\mcitedefaultendpunct}{\mcitedefaultseppunct}\relax
\EndOfBibitem
\bibitem[Kogl \latin{et~al.}(2023)Kogl, Soubelet, Brotons-Gisbert, Stier,
  Gerardot, and Finley]{Kogl23}
Kogl,~M.; Soubelet,~P.; Brotons-Gisbert,~M.; Stier,~A.~V.; Gerardot,~B.~D.;
  Finley,~J.~J. Moire straintronics: a universal platform for reconfigurable
  quantum materials. \emph{Npj 2d Mater. and Appl.} \textbf{2023}, 32\relax
\mciteBstWouldAddEndPuncttrue
\mciteSetBstMidEndSepPunct{\mcitedefaultmidpunct}
{\mcitedefaultendpunct}{\mcitedefaultseppunct}\relax
\EndOfBibitem
\bibitem[Gao \latin{et~al.}(2023)Gao, Xu, Farooq, Xian, and Huang]{Gao23}
Gao,~Y.~F.; Xu,~Q.~L.; Farooq,~M.~U.; Xian,~L.~D.; Huang,~L. Switching the
  Moire Lattice Models in the Twisted Bilayer WSe2 by Strain or Pressure.
  \emph{Nano Lett.} \textbf{2023}, \emph{23}, 7921\relax
\mciteBstWouldAddEndPuncttrue
\mciteSetBstMidEndSepPunct{\mcitedefaultmidpunct}
{\mcitedefaultendpunct}{\mcitedefaultseppunct}\relax
\EndOfBibitem
\bibitem[Balents \latin{et~al.}(2020)Balents, Dean, Efetov, and
  Young]{Balents20}
Balents,~L.; Dean,~C.~R.; Efetov,~D.~K.; Young,~A.~F. Superconductivity and
  strong correlations in moire flat bands. \emph{Nat. Phys.} \textbf{2020},
  \emph{16}, 725--733\relax
\mciteBstWouldAddEndPuncttrue
\mciteSetBstMidEndSepPunct{\mcitedefaultmidpunct}
{\mcitedefaultendpunct}{\mcitedefaultseppunct}\relax
\EndOfBibitem
\bibitem[Wu and Das~Sarma(2019)Wu, and Das~Sarma]{Wu19}
Wu,~F.~C.; Das~Sarma,~S. Identification of superconducting pairing symmetry in
  twisted bilayer graphene using in-plane magnetic field and strain.
  \emph{Phys. Rev. B} \textbf{2019}, \emph{99}, 220507\relax
\mciteBstWouldAddEndPuncttrue
\mciteSetBstMidEndSepPunct{\mcitedefaultmidpunct}
{\mcitedefaultendpunct}{\mcitedefaultseppunct}\relax
\EndOfBibitem
\bibitem[Island \latin{et~al.}(2011)Island, Tayari, Yigen, McRae, and
  Champagne]{Island11}
Island,~J.~O.; Tayari,~V.; Yigen,~S.; McRae,~A.~C.; Champagne,~A.~R.
  Ultra-short suspended single-wall carbon nanotube transistors. \emph{Appl.
  Phys. Lett.} \textbf{2011}, \emph{99}, 243106\relax
\mciteBstWouldAddEndPuncttrue
\mciteSetBstMidEndSepPunct{\mcitedefaultmidpunct}
{\mcitedefaultendpunct}{\mcitedefaultseppunct}\relax
\EndOfBibitem
\end{mcitethebibliography}


\providecommand{\latin}[1]{#1}
\makeatletter
\providecommand{\doi}
  {\begingroup\let\do\@makeother\dospecials
  \catcode`\{=1 \catcode`\}=2 \doi@aux}
\providecommand{\doi@aux}[1]{\endgroup\texttt{#1}}
\makeatother
\providecommand*\mcitethebibliography{\thebibliography}
\csname @ifundefined\endcsname{endmcitethebibliography}
  {\let\endmcitethebibliography\endthebibliography}{}
\begin{mcitethebibliography}{12}
\providecommand*\natexlab[1]{#1}
\providecommand*\mciteSetBstSublistMode[1]{}
\providecommand*\mciteSetBstMaxWidthForm[2]{}
\providecommand*\mciteBstWouldAddEndPuncttrue
  {\def\EndOfBibitem{\unskip.}}
\providecommand*\mciteBstWouldAddEndPunctfalse
  {\let\EndOfBibitem\relax}
\providecommand*\mciteSetBstMidEndSepPunct[3]{}
\providecommand*\mciteSetBstSublistLabelBeginEnd[3]{}
\providecommand*\EndOfBibitem{}
\mciteSetBstSublistMode{f}
\mciteSetBstMaxWidthForm{subitem}{(\alph{mcitesubitemcount})}
\mciteSetBstSublistLabelBeginEnd
  {\mcitemaxwidthsubitemform\space}
  {\relax}
  {\relax}

\bibitem[McRae \latin{et~al.}(2019)McRae, Wei, and Champagne]{McRae19}
McRae,~A.~C.; Wei,~G.; Champagne,~A.~R. Graphene Quantum Strain Transistors.
  \emph{Phys. Rev. Appl.} \textbf{2019}, \emph{11}, 054019\relax
\mciteBstWouldAddEndPuncttrue
\mciteSetBstMidEndSepPunct{\mcitedefaultmidpunct}
{\mcitedefaultendpunct}{\mcitedefaultseppunct}\relax
\EndOfBibitem
\bibitem[Naumis \latin{et~al.}(2017)Naumis, Barraza-Lopez, Oliva-Leyva, and
  Terrones]{Naumis17}
Naumis,~G.~G.; Barraza-Lopez,~S.; Oliva-Leyva,~M.; Terrones,~H. Electronic and
  optical properties of strained graphene and other strained 2D materials: a
  review. \emph{Rep. Prog. Phys.} \textbf{2017}, \emph{80}, 096501\relax
\mciteBstWouldAddEndPuncttrue
\mciteSetBstMidEndSepPunct{\mcitedefaultmidpunct}
{\mcitedefaultendpunct}{\mcitedefaultseppunct}\relax
\EndOfBibitem
\bibitem[Choi \latin{et~al.}(2010)Choi, Jhi, and Son]{Choi10}
Choi,~S.-M.; Jhi,~S.-H.; Son,~Y.-W. Effects of strain on electronic properties
  of graphene. \emph{Phys. Rev. B} \textbf{2010}, \emph{81}, 081407\relax
\mciteBstWouldAddEndPuncttrue
\mciteSetBstMidEndSepPunct{\mcitedefaultmidpunct}
{\mcitedefaultendpunct}{\mcitedefaultseppunct}\relax
\EndOfBibitem
\bibitem[Fogler \latin{et~al.}(2008)Fogler, Guinea, and Katsnelson]{Fogler08}
Fogler,~M.~M.; Guinea,~F.; Katsnelson,~M.~I. Pseudomagnetic fields and
  ballistic transport in a suspended graphene sheet. \emph{Phys. Rev. Lett.}
  \textbf{2008}, \emph{101}, 226804\relax
\mciteBstWouldAddEndPuncttrue
\mciteSetBstMidEndSepPunct{\mcitedefaultmidpunct}
{\mcitedefaultendpunct}{\mcitedefaultseppunct}\relax
\EndOfBibitem
\bibitem[Pellegrino \latin{et~al.}(2011)Pellegrino, Angilella, and
  Pucci]{Pellegrino11}
Pellegrino,~F. M.~D.; Angilella,~G. G.~N.; Pucci,~R. Transport properties of
  graphene across strain-induced nonuniform velocity profiles. \emph{Phys Rev
  B} \textbf{2011}, \emph{84}, 195404\relax
\mciteBstWouldAddEndPuncttrue
\mciteSetBstMidEndSepPunct{\mcitedefaultmidpunct}
{\mcitedefaultendpunct}{\mcitedefaultseppunct}\relax
\EndOfBibitem
\bibitem[Island \latin{et~al.}(2011)Island, Tayari, Yigen, McRae, and
  Champagne]{Island11}
Island,~J.~O.; Tayari,~V.; Yigen,~S.; McRae,~A.~C.; Champagne,~A.~R.
  Ultra-short suspended single-wall carbon nanotube transistors. \emph{Appl.
  Phys. Lett.} \textbf{2011}, \emph{99}, 243106\relax
\mciteBstWouldAddEndPuncttrue
\mciteSetBstMidEndSepPunct{\mcitedefaultmidpunct}
{\mcitedefaultendpunct}{\mcitedefaultseppunct}\relax
\EndOfBibitem
\bibitem[McRae \latin{et~al.}(2017)McRae, Tayari, Porter, and
  Champagne]{McRae17}
McRae,~A.~C.; Tayari,~V.; Porter,~J.~M.; Champagne,~A.~R. Giant electron-hole
  transport asymmetry in ultra-short quantum transistors. \emph{Nat. Commun.}
  \textbf{2017}, \emph{8}, 15491\relax
\mciteBstWouldAddEndPuncttrue
\mciteSetBstMidEndSepPunct{\mcitedefaultmidpunct}
{\mcitedefaultendpunct}{\mcitedefaultseppunct}\relax
\EndOfBibitem
\bibitem[Nix and MacNair(1941)Nix, and MacNair]{Nix41}
Nix,~F.~C.; MacNair,~D. The Thermal Expansion of Pure Metals: Copper, Gold,
  Aluminum, Nickel, and Iron. \emph{Phys. Rev.} \textbf{1941}, \emph{60},
  597\relax
\mciteBstWouldAddEndPuncttrue
\mciteSetBstMidEndSepPunct{\mcitedefaultmidpunct}
{\mcitedefaultendpunct}{\mcitedefaultseppunct}\relax
\EndOfBibitem
\bibitem[Yoon \latin{et~al.}(2011)Yoon, Son, and Cheong]{Yoon11}
Yoon,~D.; Son,~Y.~W.; Cheong,~H. Negative Thermal Expansion Coefficient of
  Graphene Measured by Raman Spectroscopy. \emph{Nano Lett.} \textbf{2011},
  \emph{11}, 3227\relax
\mciteBstWouldAddEndPuncttrue
\mciteSetBstMidEndSepPunct{\mcitedefaultmidpunct}
{\mcitedefaultendpunct}{\mcitedefaultseppunct}\relax
\EndOfBibitem
\bibitem[Wang \latin{et~al.}(2021)Wang, Baumgartner, Makk, Zihlmann, Varghese,
  Indolese, Watanabe, Taniguchi, and Schonenberger]{Wang21}
Wang,~L.; Baumgartner,~A.; Makk,~P.; Zihlmann,~S.; Varghese,~B.~S.;
  Indolese,~D.~I.; Watanabe,~K.; Taniguchi,~T.; Schonenberger,~C. Global
  strain-induced scalar potential in graphene devices. \emph{Commun. Phys.}
  \textbf{2021}, \emph{4}, 147\relax
\mciteBstWouldAddEndPuncttrue
\mciteSetBstMidEndSepPunct{\mcitedefaultmidpunct}
{\mcitedefaultendpunct}{\mcitedefaultseppunct}\relax
\EndOfBibitem
\bibitem[Grassano \latin{et~al.}(2020)Grassano, D'Alessandro, Pulci, Sharapov,
  Gusynin, and Varlamov]{Grassano20}
Grassano,~D.; D'Alessandro,~M.; Pulci,~O.; Sharapov,~S.~G.; Gusynin,~V.~P.;
  Varlamov,~A.~A. Work function, deformation potential, and collapse of Landau
  levels in strained graphene and silicene. \emph{Phys. Rev. B} \textbf{2020},
  \emph{101}, 245115\relax
\mciteBstWouldAddEndPuncttrue
\mciteSetBstMidEndSepPunct{\mcitedefaultmidpunct}
{\mcitedefaultendpunct}{\mcitedefaultseppunct}\relax
\EndOfBibitem
\end{mcitethebibliography}

\end{document}


\tableofcontents
\section{Applied theoretical model}\label{sec:S1}
In this section we derive step by step the applied theoretical model\cite{McRae19} used to calculate the ballistic conductance $G$ in our strained graphene devices, with the notation specific to this manuscript. The Hamiltonian in the graphene channel and contacts of our devices (see Fig. 1b) are given by Eq. \ref{Eq:Hamiltonian_chan} and Eq. \ref{Eq:Hamiltonian_cont}, respectively.
\begin{equation}
\label{Eq:Hamiltonian_chan}
H_{K_{i}, channel}=\hbar v_{F}(\bar{\bm{I}}+(1-\beta)\bar{\bm{\varepsilon}})\cdot\bm{\sigma}\cdot(\bm{k}-\bm{A}_{i}) + \Delta\mu_{\text{G}} + e\phi_{\varepsilon},
\end{equation}
\begin{equation}
\label{Eq:Hamiltonian_cont}
H_{K_{i},contact}=\hbar v_{F}\bm{\sigma}\cdot\bm{k} \text{ } + \mu_{\text{contact}},
\end{equation}
The pseudospin operator $\bm{\sigma}=(\sigma_x,\sigma_y)$ is represented by the Pauli matrices and acts on the two-component spinor wavefunction referring to the A and B sublattices. The pseudospin orientation is either parallel (up) or anti-parallel (down) with the generalized wave vector $\bm{k} - \bm{A}_{i}$, where the index $i =$ 1, 2, 3 labels the three $K$ valleys (Fig. \ref{fig.s1}). The matrices $\bar{\bm{I}}$ and $\bar{\bm{\varepsilon}}$ are respectively the identity matrix and strain tensor. In the device's $x$ - $y$ coordinates, $\bar{\bm{\varepsilon}}$ has elements $\varepsilon_{xx}=\varepsilon_{\text{total}}$, $\varepsilon_{yy}=-\nu \varepsilon_{\text{total}}$ and $\varepsilon_{xy}=\varepsilon_{yx}=0$, where $\nu=0.165$ is the Poisson ratio \cite{Naumis17}. The term $\Delta\mu_{\text{G}}=\hbar v_F\sqrt{\pi n_{total}}$  is the gate-induced electrostatic potential in the channel, where $n_{tot} = \sqrt{((c_{\text{G}}/e) (V_{\text{G}}-V_{\text{D}}))^2 + n_{rms}^2 }$, $n_{rms}$ is the impurity induced minimal channel doping, $v_{F}=$ 1.0 $\times$ 10$^6$ m/s is the Fermi velocity, and $c_{\text{G}}$ is the gate-channel capacitance per unit area.

As per Eq. \ref{Eq:Hamiltonian_chan}, uniaxial strain has three main qualitative effects on the channel's band structure. First, a downward shift of the Fermi energy which can be described by a scalar potential $e\phi_{\varepsilon} = g_{\varepsilon}(1-\nu)\varepsilon_{\text{total}}$, where\cite{Choi10} $g_{\varepsilon}\approx 3.0$~eV. Secondly, the position of the Dirac points shift in momentum space, and these shifts can be described by gauge vector potentials $\bm{A}_{i}$. Thirdly, there is an anisotropic distorsion the Dirac cones which corresponds to a direction-dependent Fermi velocity $\bar{\bm{v}}_{F} =  v_{F}(\bar{\bm{I}}+(1-\beta)\bar{\bm{\varepsilon}})$. For uniaxial strain, $\bar{\bm{v}}_{F}$ has only diagonal elements $v_{xx}=1+(1-\beta)\varepsilon_{\text{total}}$ and $v_{yy}=1-(1-\beta)\nu\varepsilon_{\text{total}}$. The parameter $\beta \approx 2.5$ is the electronic Gr\"{u}neisen parameter \cite{Naumis17}.
The Hamiltonian in the source/drain graphene contacts (Eq. \ref{Eq:Hamiltonian_cont}) has no strain-induced terms and $\Delta\mu_{\text{G}}$ is replaced with $\mu_{\text{contact}}$, the fixed graphene contact doping due to charge transfer from the gold film.
We now derive the charge conductance across our Devices. First, we will solve the Dirac equation in the presence of a uniaxial strain to find the eigenstates of the charge carriers. Then we will use the boundary conditions to solve for the transmission amplitude (and probability). Finally, we will sum the transmission of each conduction mode to obtain the device's conductance.
\subsection{Solutions of the Dirac equation with uniaxial strain}
In our strained graphene channels, the Dirac equation can be written as \cite{McRae19}:
\begin{equation}
\label{Eq:Dirac_Eq_K_Hat_Str}
\hbar v_F
\begin{pmatrix}
\sigma_x, \sigma_y
\end{pmatrix}.
(\bar{I}+(1-\beta)\bar{\varepsilon}).
\begin{pmatrix}
\tilde{k_{x}} \\
\tilde{k_{y}}
\end{pmatrix}
\Psi
= \tilde{E_n} \Psi
\end{equation}
where $\bar{I}$ is the identity matrix, $\bar{\varepsilon}$ is the strain tensor, $\beta$ is the electronic Grüneisen parameter. The wave function $\Psi$ has the following form:
\begin{equation} \label{Eq:Wavef_K_pq_Z_Str}
\Psi_{\tilde{k_y},y,\tilde{k_x},x} = a_n
\begin{pmatrix}
1\\
z_{\tilde{k_y},\tilde{k_x}}
\end{pmatrix}
e^{i\tilde{k_y}y} e^{i\tilde{k_x}x}
\end{equation}
Then Eq. \ref{Eq:Dirac_Eq_K_Hat_Str} becomes
\begin{equation}
\label{Eq:Dirac_Eq_K_Z_Str}
\begin{pmatrix}
(\tilde{k_x}v_{xx} - i\tilde{k_y}v_{yy}) z_{\tilde{k_y},\tilde{k_x}}\\
\tilde{k_x}v_{xx} + i\tilde{k_y}v_{yy}
\end{pmatrix}
=
\begin{pmatrix}
\pm \sqrt{(\tilde{k_x}v_{xx})^2 + (\tilde{k_y}v_{yy})^2}\\
\pm \sqrt{(\tilde{k_x}v_{xx})^2 + (\tilde{k_y}v_{yy})^2}z_{\tilde{k_y},\tilde{k_x}}
\end{pmatrix}
\end{equation}
Thus,
\begin{equation} \label{Eq:Z_K_pnE_Str}
z_{\tilde{k_y},\tilde{k_x}} = \text{sgn}(\tilde{k_F}) \frac{\tilde{k_x}(v_{xx}) + i\tilde{k_y}(v_{yy})}{\sqrt{(\tilde{k_x}v_{xx})^2 + (\tilde{k_y}v_{yy})^2}}
\end{equation}
where $\tilde{k_F} = \tilde{E_n}/(\hbar v_F)$. We note the identity,
\begin{equation} \label{Eq:Z_K_Iden_Str}
z_{\tilde{k_y},\tilde{k_x}}z_{\tilde{k_y},-\tilde{k_x}} = -1
\end{equation}
Using Eq. \ref{Eq:Z_K_pnE_Str} in Eq. \ref{Eq:Wavef_K_pq_Z_Str}, the plane wave solutions of the Dirac equation in strained graphene are written as:
\begin{equation} \label{Eq:Wavef_K_pq_Str}
\Psi_{\tilde{k_y},y,\tilde{k_x},x} = a_n
\begin{pmatrix}
1\\
\text{sgn}(\tilde{k_F}) \frac{\tilde{k_x}v_{xx} + i\tilde{k_y}v_{yy}}{\sqrt{(\tilde{k_x}v_{xx})^2 + (\tilde{k_y}v_{yy})^2}}
\end{pmatrix}
e^{i\tilde{k_y}y} e^{i\tilde{k_x}x}
\end{equation}
\subsection{Transmission amplitude and conductance in strained graphene}
\begin{figure}[htbp]
  \centering
\includegraphics[scale=1]{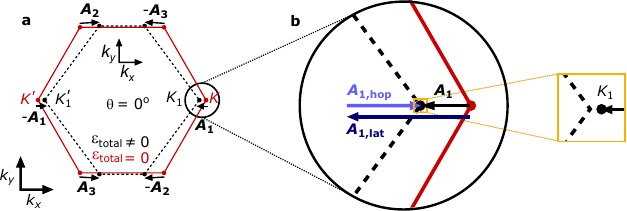}
\caption{First Brillouin zone (FBZ) of graphene under uniaxial strain. \textbf{a} Unstrained (red) and uniaxially strained (black) FBZ of graphene when $\theta=0^{\circ}$. The strain value in this figure is exaggerated, $\varepsilon_{\text{tot}}=20$ \%, to make its effects clearly visible. \textbf{b} Under strain, the Dirac point shifts define the gauge vector potentials (blue arrows), $\bm{A}_{i} = \bm{A}_{\text{lat},i} + \bm{A}_{\text{hop}}$. The outset shows that the corner of the FBZ does not coincide with the Dirac point under strain.}
\label{fig.s1}
\end{figure}

Based on the cartoon of the device in Fig. 1b of the main text, we label the source contact as $S$ and drain contact as $D$. We use $\tilde{\bm{k}}$ and $\bm{k}$ to represent the wavevectors in the channel and contacts, respectively. We now write the total wavefunction, including the reflected and transmitted $\Psi$ in the three regions of the device for an incident energy $E_{F} = \mu_{contact}$ and for a carrier in the $n$-th transverse momentum mode (subband) $k_y = (\pi/W)(n + 1/2)$, for $n\geq$ 0, and $k_y = (\pi/W)(n - 1/2)$, for $n\leq$ 0.
\begin{align}
	&\Psi =
		\begin{cases}
			\Phi_S, & \text{if} \ x < 0 \\
			\tilde{\Phi}, & \text{if} \ 0 < x < L \\
			\Phi_D, & \text{if} \ x > L
		\end{cases} \\
	&\Phi_S = \Psi_{k_y,y,k_x,x} + r_n \Psi_{k_y,y,-k_x,x} \\
	&\tilde{\Phi} = \alpha_n \Psi_{\tilde{k_y},y,\tilde{k_x},x} + \beta_n \Psi_{\tilde{k_y},y,-\tilde{k_x},x} \\
	&\Phi_D = t_n\Psi_{k_y,y,k_x,x-L}
\end{align}
where $r_n$, $t_n$ are the reflection and transmission amplitudes, and $\alpha_n$, $\beta_n$ are coefficients. Because of the continuity of $\Psi$ at $x = 0$ (source-channel edge) and $x = L$ (channel-drain edge), we obtain the following equations.
\begin{align}
\label{Eq:Cont_x_0_Str}
&\Psi_{k_y,y,k_x,0} + r_n \Psi_{k_y,y,-k_x,0} = \alpha_n \Psi_{\tilde{k_y},y,\tilde{k_x},0} + \beta_n \Psi_{\tilde{k_y},y,-\tilde{k_x},0} \\
\label{Eq:Cont_x_L_Str}
&\alpha_n \Psi_{\tilde{k_y},y,\tilde{k_x},L} + \beta_n \Psi_{\tilde{k_y},y,-\tilde{k_x},L} = t_n\Psi_{k_y,y,k_x,L-L}
\end{align}
By solving Eq. \ref{Eq:Cont_x_0_Str} and \ref{Eq:Cont_x_L_Str}, one finds the transmission amplitude:
\begin{equation} \label{Eq:t_n_z_Str}
t_n = \frac{(1 + z_{k_y,k_x}^2)(1+z_{\tilde{k_y},\tilde{k_x}}^2)}{e^{i\tilde{k_x}L}(z_{k_y,k_x}-z_{\tilde{k_y},\tilde{k_x}})^2 + e^{-i\tilde{k_x}L}(1+z_{k_y,k_x}z_{\tilde{k_y},\tilde{k_x}})^2}
\end{equation}
where we have used the identity Eq. \ref{Eq:Z_K_Iden_Str}. After the substitution for $z_{k_y,k_x}, z_{\tilde{k_y},\tilde{k_x}}$,
\begin{equation} \label{Eq:t_n_Str}
\resizebox{0.98\hsize}{!}{%
$t_n = \frac{k_x\tilde{k_x}v_{xx}}{k_x \tilde{k_x}v_{xx} \cos(\tilde{k_x}L)-i\big(-k_y\tilde{k_y}v_{yy}+\text{sgn}(k_F\tilde{k_F})\sqrt{(k_{x}^2+k_{y}^2)((\tilde{k_x}v_{xx})^2+(\tilde{k_y}v_{yy})^2)}\big)\sin(\tilde{k_x}L)}$%
}
\end{equation}
The transmission probability of $n$-th mode equals $|t_n|^2$,
\begin{equation} \label{Eq:T_n_Str_1}
\resizebox{0.98\hsize}{!}{%
$T_n = \frac{k_{x}^2\tilde{k_x}^2(v_{xx})^2}{k_{x}^2\tilde{k_x}^2(v_{xx})^2\cos^2(\tilde{k_x}L)+\big(-k_y\tilde{k_y}v_{yy} +\text{sgn}(k_F\tilde{k_F})\sqrt{k_{x}^2+k_{y}^2}\sqrt{\tilde{k_x}^2(v_{xx})^2+\tilde{k_y}^2(v_{yy})^2}\big)^2\sin^2(\tilde{k_x}L)}$%
}
\end{equation}
where $\tilde{k}_y = k_y - A_y$.
Although the notation differs slightly, and additional terms are included (corrections to $v_{F}$ and all $A_{y}$ contributions as per \ref{Eq:VectorPots} below), the result in Eq. \ref{Eq:T_n_Str_1} is fully consistent with previous work \cite{Fogler08,Pellegrino11,McRae19}. The strain-induced changes to the tight-binding nearest-neighbour hopping amplitude and the nearest-neighbour inter-atomic distances are both incorporated in the Dirac Hamiltonian of graphene as gauge vector potentials $\bm{A}_{i}$. These total gauge potentials are $\bm{A}_{i}=\bm{A}_{\text{lat},i}+\bm{A}_{\text{hop}}$, and are given in \ref{Eq:VectorPots} and shown in \ref{fig.s1}.
\begin{subequations} \label{Eq:VectorPots}
\begin{eqnarray}
&\bm{A}_{\text{hop}}&=\frac{\beta\varepsilon(1+\nu)}{2a} \begin{pmatrix}
 \cos3\theta\\ \sin3\theta \end{pmatrix}
  \\[10pt]
&\bm{A}_{\text{lat,1}}&=\frac{4\pi\varepsilon}{3\sqrt{3}a} \begin{pmatrix}
 -\cos\theta\\ \nu\sin\theta \end{pmatrix}
 \\[10pt]
&\bm{A}_{\text{lat,2}}&=\frac{2\pi\varepsilon}{3a} \begin{pmatrix}
 \frac{1}{\sqrt{3}}\cos\theta +\sin\theta \\ -\frac{1}{\sqrt{3}}\nu\sin\theta +\nu\cos\theta \end{pmatrix}
 \\[10pt]
&\bm{A}_{\text{lat,3}}&=\frac{2\pi\varepsilon}{3a} \begin{pmatrix}
 \frac{1}{\sqrt{3}}\cos\theta -\sin\theta \\ -\frac{1}{\sqrt{3}}\nu\sin\theta -\nu\cos\theta \end{pmatrix}
\end{eqnarray}
\end{subequations}
The charge carriers' trajectories are described by their momentum wavevectors in the source, $\bm{k} = \pm(k_x\hat{x} + k_y\hat{y})$, and in the channel, $\bm{\tilde{k}} = \pm(\tilde{k_x} \hat{x} + \tilde{k_y} \hat{y})$, where the $\pm$ symbol refers to electron or hole transport, respectively. The $y$-axis boundary condition conserves the total $y$-momentum throughout the device, such that $\tilde{k_y}=k_y - A_{i,y}$. This strain-induced shift in $\tilde{k_y}$ alters the propagation angle of the carriers in the channel, and thus their Klein transmission probability at the strained/unstrained interfaces. We can simplify Eq. \ref{Eq:T_n_Str_1}, by using $\tilde{k}_y = k_y - A_y$,  $k_y=\frac{\pi}{W}(n+\tfrac{1}{2})$, $k_x=({k_{F}^2-k_{y}^2})^{1/2}$, and $\tilde{k_x}=v_{xx}^{-1}[{\tilde{k}_F^2-v_{yy}^2(k_{y}-\xi A_{i,y})^2}]^{1/2}$. We now write $T_{\xi,i,n}$ in mode $n$ and valley $\xi = \pm 1, i = 1,2,3$, as
\begin{equation} \label{Eq:StrainTrans}
T_{\xi,i,n}=\frac{(v_{xx}k_x\tilde{k_x})^2}{(v_{xx}k_x\tilde{k_x})^2\cos^2(\tilde{k_x}L) + (k_{F}\tilde{k}_{F}-v_{yy}k_y(k_y-\xi A_{i,y}))^2 \sin^2(\tilde{k_x}L)}
\end{equation}
Finally, we calculate the conductance of the device by properly summing the transmissions from all relevant modes:
\begin{equation}\label{Eq:StrainCond}
G=\frac{2e^2}{h}\frac{1}{3} \sum_\xi\sum_i^3\sum_n^N T_{\xi,i,n}
\end{equation}
where $N=\text{Int}(k_F W/\pi +1/2)$ is the number of energetically allowed modes set by the contact's Fermi energy, $\mu_{contact}$, and the factor $\frac{1}{3}$ accounts for the lifting of the three-fold $K$ and $K'$ point degeneracy in strained graphene. We use Eq.\ \ref{Eq:StrainCond} to calculate the theoretical $G_{channel}$ values presented in the main text and following sections.

\section{Additional information on instrumentation}
\begin{figure}
\includegraphics[scale=0.94]{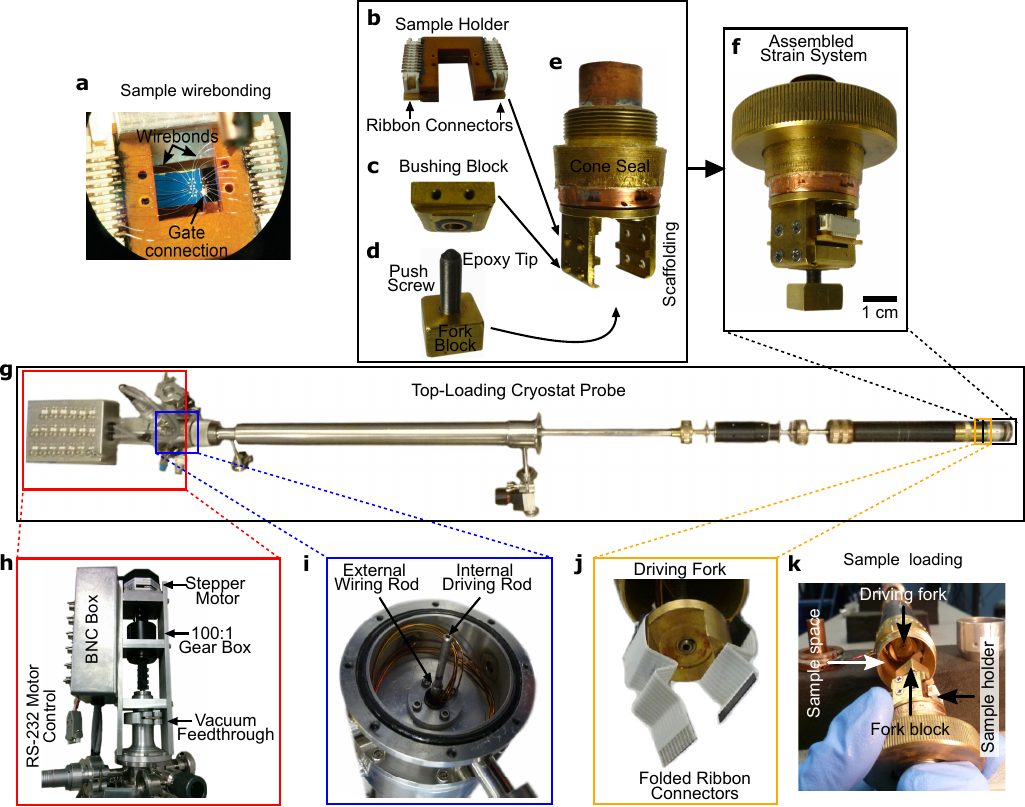}
\caption{Custom-built electro-mechanical quantum transport instrumentation. \textbf{a} Top-view of a Si chip mounted on the sample holder shown in \textbf{b}. Multiple devices are nanofabricated on one chip and wire-bonded using a manual wire bonder. \textbf{c} Bushing used to anchor the push screw shown in \textbf{d}. \textbf{e} The cone seal on which the screw and sample assembly are mounted as shown in \textbf{f}. The cone seal mounts at the end of the top-loading cryostat probe in \textbf{g}. At the top of the probe there is a stepper motor assembly shown in \textbf{h}, which is connected to a internal driving rod in \textbf{i}. It transmits the motion down the cryostat to a driving fork shown in \textbf{j}, which rotates the pushing screw via the fork block in \textbf{d}. The ribbon connectors visible in \textbf{j} are connected to the sides of the sample holder in \textbf{b}. The panel \textbf{k} shows how a sample is loaded into the cryostat.}
\label{fig.s2}
\end{figure}

Figure \ref{fig.s2} shows additional components of our instrumentation, how some key components are assembled, and how the samples are mounted before cooling down.
\clearpage
\section{Additional information on strain calibration}
\subsection{Tunnel junction J2}
\begin{figure}
  \centering
\includegraphics[scale=1.5]{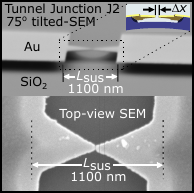}
\caption{Tunnel Junction J2. The top image shows a tilted-SEM image of the device where the total length of the suspended gold cantilever arms, $L_{sus}$ = 1100 $\pm$ 50 nm, is visible. The bottom image is a top-view of the same device.}
\label{fig.s3}
\end{figure}

The scanning electron microscope (SEM) images of Device J2 (tunnel junction) are shown in Fig. \ref{fig.s3}, and its data are shown in Fig. 2c of the main text. We see that the suspension length is 1100 nm $\pm$ 50 nm, both from a side view (tilted) image and a top-view image. The absence of a substrate's thermal anchoring leads to higher temperatures in the suspended gold beam during current annealing. The top-view image shows a different contrast for the gold film which was annealed, corresponding to the suspended portion of the gold.
\subsection{Notes on the calibration of the screw displacement and mechanical strain}
The theoretical (geometric) relationship between the displacement of the push screw ($\Delta z$) and the longitudinal channel elongation ($\Delta x$) is :
\begin{equation}
  \label{eq:theoretical dx vs dz}
  \Delta x=\frac{3ut}{D^2}\Delta z
\end{equation}
Where $D=8.18$ mm is the distance between the Si substrate's clamps, $t=$ 200 $\mu$m is the substrate's thickness, and $u=$ 500 $-$ 1100 nm is the total suspension length of the device (channel and contacts). The experimental calibration of $\Delta x$ is done in two independent ways. First by measuring the resistance of tunnel junctions, and secondly from Dirac point shifts $\Delta V_{\text{D}}$ in suspended graphene channels. Both methods systematically agree with each other and are within 10 \% of the geometric expectation. Most importantly, the mechanical motion of the reported devices is extremely stable and reproducible, with a $\Delta x$ resolution of the order of a few pm (10$^{-12}$ m). The first calibration is done with tunnel junctions as shown in Fig. 2 of the main text. Precisely controlled electromigration \cite{Island11,McRae17} creates nm-sized gaps in the center of suspended bow-tie-shaped gold bridges. When its tunnel gap is stretched, the tunnel junction resistance changes as:
 \begin{equation}
  \label{eq:tunnel junction resistance}
  R \propto e^{2\kappa \Delta x}, \kappa=\frac{\sqrt{2m_e \phi_{Au}}}{\hbar}=1.18\times10^{10} \mathrm{m}^{-1}
 \end{equation}
Where $m_e=9.109\times10^{-31}$ kg is the electron mass, and $\phi_{Au}\approx5.3$ eV is the work function of gold. We fit log$_{\text{10}}(R)$ vs $\Delta z$ as shown in Fig. 2b-c to obtain a calibration of $\Delta x /\Delta z= $ 9.0 $\pm$ 1 $\times$10$^{-6}$. We focus on very resistive junctions, $R>$ 10 G$\Omega$, such that the gap between the two gold cantilevers is large on an atomic scale. This avoids fluctuations in $R$ due to the roughness of our tunnel junction ``tips".

The second evidence showing that we understand quantitatively the stretching ($\Delta x$) of graphene, makes use of an intrinsic strain gauge built into graphene. Stretching graphene changes its second-nearest neighbor lattice distances, which creates a scalar potential shifting energy of its the Dirac point \cite{Choi10}. This is shown in Fig. 3 of the main text. We find that the magnitude of the gate voltage Dirac point shifts $\Delta V_{\text{D}}$ are in good agreement with the theoretically predicted ones (using the strain values from our tunnel junction calibration). Moreover, we find that $\Delta V_{\text{D}}$ in all our four samples (T1, T2, Device 1 and Device 2) are consistent.

\subsection{Notes on the thermal strain calibration}
The strain-induced shift of the Dirac point, $\Delta V_{\text{D}}$, (i.e. shift of the conductance minimum location in $G$-$V_{G}$ data) can also be used to calibrate the thermally-induced strain in our graphene channels. The expected shift induced by strain is:
\begin{equation}
  \label{eq:dirac point shift tot}
\Delta V_{\text{D}}=-\frac{e}{c_G}\frac{g^2_\varepsilon}{\pi(\hbar v_F)^2}(1-\nu)^2\varepsilon\textsubscript{tot}^2 = -\frac{e}{c_G}\frac{g^2_\varepsilon}{\pi(\hbar v_F)^2}(1-\nu)^2\varepsilon\textsubscript{thermal}^2
\end{equation}
Since $\varepsilon\textsubscript{tot}= \varepsilon\textsubscript{thermal} + \varepsilon\textsubscript{mech}$, and we conducted our thermal strain calibration at $\varepsilon\textsubscript{mech} =$ 0, we have $\varepsilon\textsubscript{tot}=\varepsilon\textsubscript{thermal}$.
There parameter $\nu$ is the Poisson ratio 0.165, $g_\varepsilon$ is the scalar potential prefactor which we seek to measure, and $c_G$ is the capacitance per unit area which given by:
\begin{equation}
  \label{eq:gate capacitance per unit area}
c_G=\frac{\epsilon_{vac}\epsilon_{ox} }{t_{ox}\epsilon_{vac}+t_{vac}\epsilon_{ox}}
\end{equation}
Where $t_{ox}$ and $t_{vac}$ are respectively the thickness of the SiO$_{2}$ layer and vacuum layer between the Si back-gate and graphene channel. The values of $c_G$ in our four reported device are respectively: 4.77 (Device T1), 3.60 (Device T2), 3.72 (Device 1), 3.34 (Device 2) $\times$10$^{-5}$ F/(m$^2$).
We calculate the expected thermal strain $\varepsilon_{\text{thermal}}$ in the graphene channel at temperature $T$ using the thermal contraction/expansion of gold and graphene as follows:
\begin{equation}\label{eq:ThermalStrain}
\varepsilon_{\text{thermal}}=  - \frac{u-L}{L} \int_{300}^T \alpha_{\text{Au}}(t)dt -\int_{300}^T \alpha_{\text{g}}(t)dt
\end{equation}
where $\alpha_{\text{Au}}$ and $\alpha_{\text{g}}$ are the coefficients of thermal expansion for gold and graphene respectively.  Using  $\alpha_{\text{Au}}$ from \textit{Nix et al.}\cite{Nix41} and $\alpha_{\text{g}}$ from \textit{Yoon et al.}\cite{Yoon11}, we calculate  a net thermal strain of $\varepsilon_{\text{thermal}}=$ 1.55 \% and 1.18 $\%$ at $T=1.3$ K for Devices 1 and 2, respectively. The systematic uncertainty on $\varepsilon_{\text{thermal}}$ is $\approx$ 0.1 \%, and stems from the uncertainty on $L_{\text{sus}}$. The numbers for Devices T1 and T2 at various temperatures are shown in Fig. 2f of the main text.
We used Eq. \ref{eq:dirac point shift tot}, to fit the data shown in Fig. 2f and extracted $g_{\varepsilon} =$ 3.03 and 2.67 in Devices T1 and T2. These values are in close agreement with the theoretically and experimentally reported values \cite{Wang21,Grassano20,Pellegrino11,Choi10}, and confirm that we have a good understanding of the thermal strain in our devices. We find very similar values for $g_{\varepsilon}$ in Devices 1 and 2, where we used $\varepsilon_{\text{mech}}$ (Figs. 3e and 5g) instead of $\varepsilon_{\text{thermal}}$ to induce $\Delta V_{\text{D}}$.

\section{Additional mechanical data sweeps for Device 1}
\begin{figure}[hbtp]
  \centering
\includegraphics[scale=1]{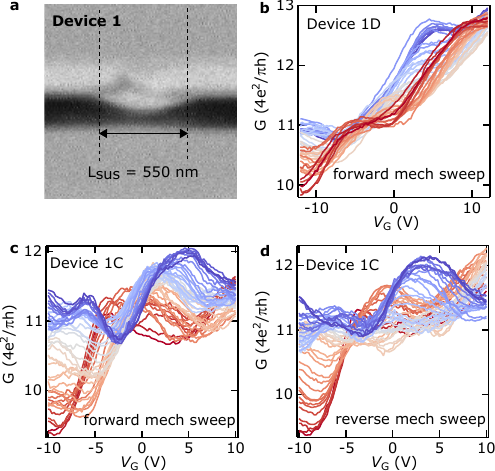}
\caption{Additional raw data for Device 1. \textbf{a} Tilted-SEM image of Device 1 showing the suspension length. \textbf{b}-\textbf{c}-\textbf{d} $G$ - $V_{\text{G}}$ data for Device 1D (forward mechanical sweep), Device 1C (forward), and Device 1C (reverse), respectively. The exact strain ranges are shown in Fig. 3e, and are $\sim$ 1 \%.}
\label{fig.s4}
\end{figure}

Figure \ref{fig.s4} shows additional data to support the data in Fig. 3 of the main text.
Figure \ref{fig.s4}a shows a tilted-SEM image of $L_{sus}$ in Device 1. While Fig. \ref{fig.s4}b-d shows the raw data $G$ - $V_{\text{G}}$ in Device 1D (forward mechanical sweep), Device 1C (forward), and Device 1C (reverse). Note that the reverse sweep for Device 1D was not recorded during the experiment. By fitting the $V_{\text{G}}$ location of the conductance minima in these data traces from the lowest strain (blue data) to the highest strain (red data) we extracted the data points shown in Fig. 3e of the main text.

\section{Device 2: dimensions, annealing, scalar potential} 
\begin{figure}
\centering
\includegraphics[scale=1]{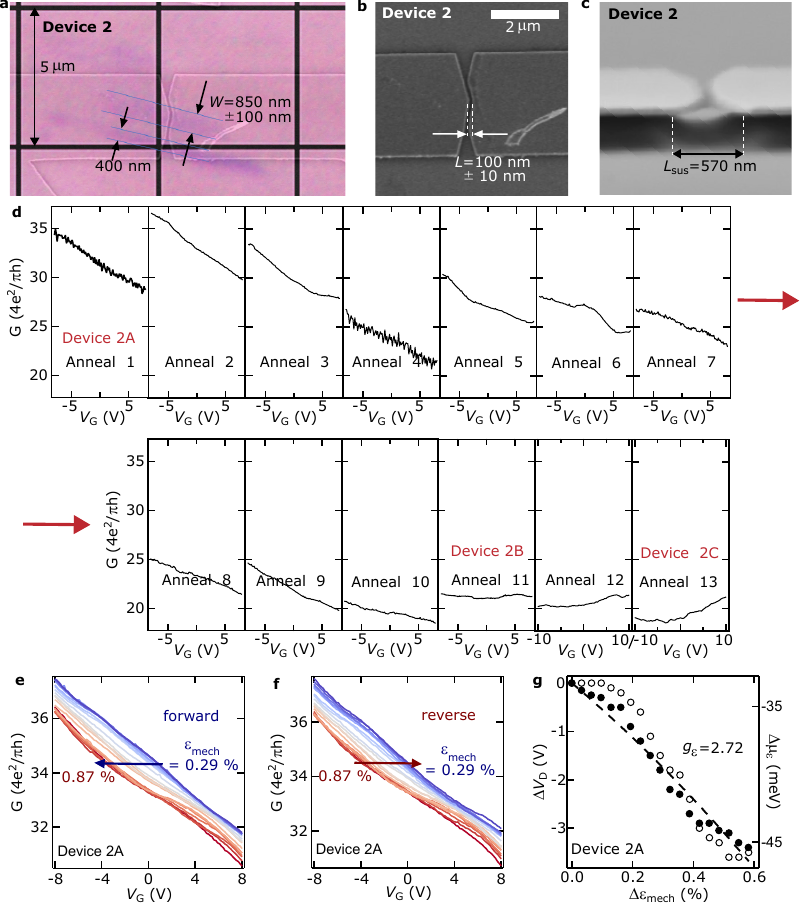}
\caption{Device 2 images, annealing data, and mechanical scalar potential data. \textbf{a} Superposition of an optical image of the Device's graphene crystal and SEM image of the lithographically defined gold clamps, showing $W=$ 850 $\pm$ 100 nm for the channel. \textbf{b} SEM image of the gold clamps showing $L=$ 100 $\pm$ 10 nm for the channel. \textbf{c} Titled-SEM image showing $L_{sus}=$ 570 $\pm$ 50 nm. \textbf{d} $G$ - $V_{\text{G}}$ data after each annealing step, at increasing power, for Device 2. The channel evolves from p-doped in Anneal 1 to n-doped in Anneal 13. \textbf{e} Forward mechanical sweep for Device 2A, showing $G$ - $V_{\text{G}}$ at each mechanical step. \textbf{f} Reverse mechanical sweep for Device 2A, showing $G$ - $V_{\text{G}}$ at each mechanical step. \textbf{g} The relative shift, $\Delta V_{\text{D}}$, of each trace in \textbf{e} and \textbf{f} versus $\Delta \varepsilon_{\text{mech}}$. The dashed line is a theoretical fit using Eq. \ref{eq:dirac point shift tot}.}
\label{fig.s5}
\end{figure}

Figure \ref{fig.s5} shows additional data from Device 2 to support the data in Fig. 3. In Fig.\ \ref{fig.s5}a, we see in dark purple the graphene crystal forming both the graphene contacts (under the gold clamps) and the naked suspended channel. The width of the channel is $W=$ 850 $\pm$ 100 nm, and the length is shown in Fig.\ \ref{fig.s5}b to be $L=$ 100 $\pm$ 100 nm. We note that there is secondary channel, much narrower and much longer, in parallel to the main channel in Device 2. However, because $G_{\text{channel}} \propto W/L$ and this aspect ratio for the secondary channel is a factor of 10 smaller than for the main channel, its conductance makes a very small contribution ($\sim$ 10 \%) to Device 2. Since our leading source of uncertainty is larger than 10 \%, i.e. errors in channel dimensions, we neglect the effect of this secondary channel in our data analysis. We also note that the length of the secondary channel is almost 5 times longer than the main channel, which reduces by the same factor the applied strain and the strain-induced changes in this secondary channel.  

\section{Extracting $\theta$, $\mu_{contact}$, and $n_{rms}$ in Device 1}
\begin{figure}[hbtp]
  \centering
\includegraphics[scale=1]{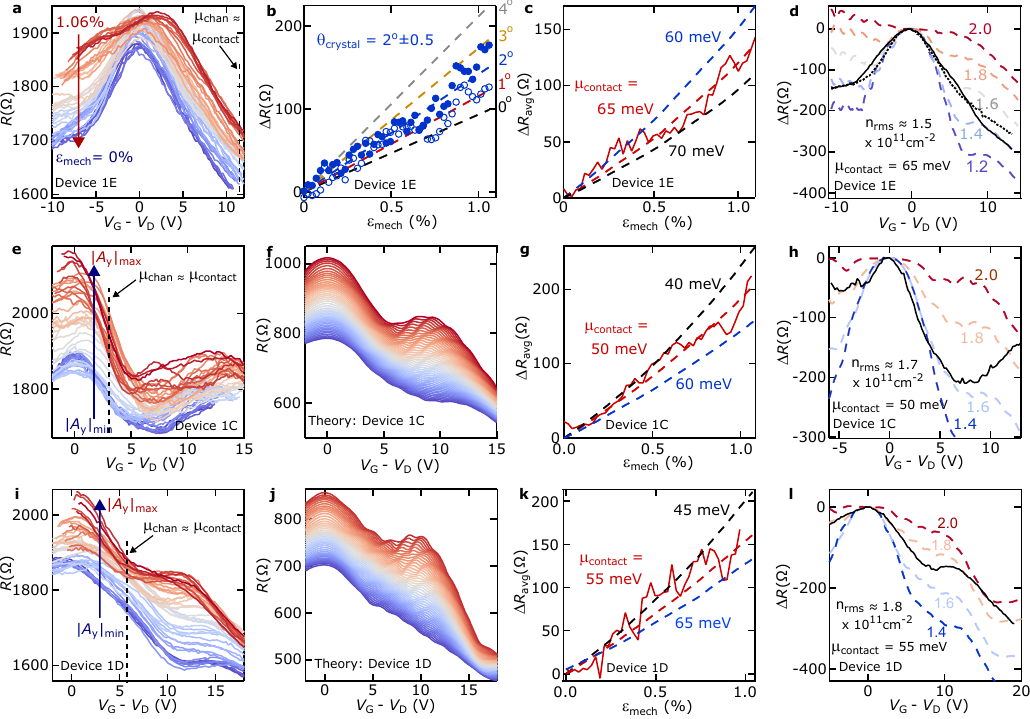}
\caption{Extracting $\theta$, $\mu_{\text{contact}}$, and $n_{\text{rms}}$ in Device 1. \textbf{a} Reverse mechanical sweep in Device 1E. \textbf{b} Forward/reverse data (solid/open circles) extracted at $V_{\text{G}}-V_{\text{D}}=$ 10.3 V, near the vertical dashed line in Fig. 4a and in \textbf{a}. The dashed lines are calculations based on Eq. 2 in the main text for various values of $\theta$. \textbf{c} The solid trace is the average change in $R$ at all $V_{\text{G}}-V_{\text{D}}$ values, $\Delta R_{\text{avg}}$, versus $\varepsilon_{\text{mech}}$ from Fig.\ 4a and in \textbf{a}. The dashed lines are $\Delta R_{\text{avg}}$ for the theoretical calculations using $\theta = $ 2.0$^o$ and varying $\mu_{\text{contact}}$. \textbf{d} Solid/dotted black data traces from the forward/reverse mechanical sweeps for Device E at the mid-range value of $\varepsilon_{\text{mech}}$. The dashed lines are calculations using $\theta = $ 2.0$^o$, $\mu_{\text{contact}} =$ 65 meV, and varying $n_{\text{rms}}$. \textbf{e, f, g, h} Forward mechanical sweep data for Device 1C, theoretical calculations, extraction of $\mu_{\text{contact}}$, and $n_{\text{rms}}$, respectively. \textbf{i, j, k, l} Forward mechanical sweep data for Device 1D, theoretical calculations, extraction of $\mu_{\text{contact}}$, and $n_{\text{rms}}$, respectively.}
\label{fig.s6}
\end{figure}

Figure \ref{fig.s6} shows the data analysis steps and calculations for Device 1's data, to extract the parameters $\theta$ (crystal orientation with respect to the strain direction), $\mu_{\text{contact}}$ (Fermi level in the contacts), and $n_{\text{rms}}$ (minimum charge doping of the channel due to random impurities).
\\
Figure \ref{fig.s6}a shows the reverse mechanical sweep in Device 1E to complement the forward data shown in Fig. 4a. Fig. \ref{fig.s6}b shows the extracted data points along the vertical cuts at $V_{\text{G}}-V_{\text{D}}=$ 10.3 V in Fig. \ref{fig.s6}a (open markers) and Fig. 4a (solid markers). These data are compared to calculations (dashed lines) based on Eq.\ 2 in the main text for various $\theta$. We find that $\theta = $ 2.0$^o$ $\pm$  0.5$^{o}$ in Device 1. We then use this quantity to model Devices 1E, 1D and 1C since the crystal orientation is not modified by Joule annealing. Fig. \ref{fig.s6}c shows data (solid trace) representing the averaged change in $R$ at all $V_{\text{G}}-V_{\text{D}}$ values, $\Delta R_{\text{avg}}$, from Figs. 4a and \ref{fig.s6}a versus $\varepsilon_{\text{mech}}$. The dashed lines are the $\Delta R_{\text{avg}}$ from the theoretical model using $\theta = $ 2.0$^o$ and varying $\mu_{\text{contact}}$. As expected, $\Delta R_{\text{avg}}$ decreases monotonously with increasing $\mu_{\text{contact}}$, and the data-theory comparison allows us to extract the experimental $\mu_{\text{contact}} =$ 65 $\pm$ 5 meV for Device 1E.

Panel d shows data (solid/dotted lines for the forward/reverse sweeps) for Device 1E at mid-range value of $\varepsilon_{\text{mech}}$. Superposed on this data are the theory calculations (dashed lines) using $\theta = $ 2.0$^o$ and $\mu_{\text{contact}} =$ 65 meV, for various $n_{\text{rms}}$. We observe as expected that the width of the conductance minimum increases with increasing $n_{\text{rms}}$ and allows us to estimate $n_{\text{rms}}=$ 1.5 $\pm$ 0.1 $\times$ 10$^{11}$ cm$^{-2}$ in Device 1E. We repeat the same steps in Figs. \ref{fig.s6}e-h to show the forward mechanical sweep data from Device 1C, corresponding theoretical calculations, extraction of $\mu_{\text{contact}}=$ 50 $\pm$ 5 meV, and $n_{rms}=$ 1.7 $\pm$ 0.1 $\times$ 10$^{11}$ cm$^{-2}$, respectively. Finally, in Figs. \ref{fig.s6}i-l, we show the forward mechanical sweep data for Device 1D, corresponding theoretical calculations, extraction of $\mu_{\text{contact}}=$ 55 $\pm$ 5 meV, and $n_{\text{rms}}=$ 1.8 $\pm$ 0.1 $\times$ 10$^{11}$ cm$^{-2}$, respectively.

\section{Evidence for mechanical control of $G$ in Device 2}
\begin{figure}[hbtp]
  \centering
\includegraphics[scale=1]{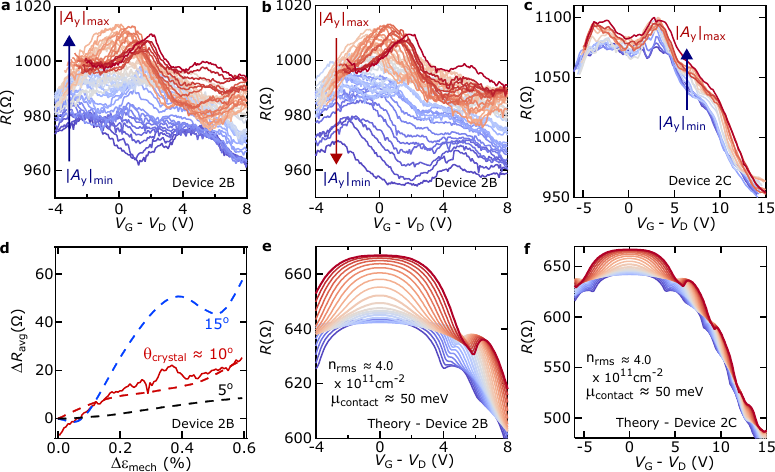}
\caption{Evidence for mechanical control of ballistic conductance in Device 2. \textbf{a} and \textbf{b} are the forward and reverse mechanical $R$ - ($V_{\text{G}}-V_{\text{D}}$) data in Device 2B (Fig. \ref{fig.s5}d) for $\varepsilon_{\text{mech}}$ ranging from 0.29 \% (dark blue trace) up to 0.87 \% (dark red trace). \textbf{c} $R$ - ($V_{\text{G}}-V_{\text{D}}$) data in Device 2C (Fig. \ref{fig.s5}d) over a range of $\varepsilon_{\text{mech}}$ from 0.28 \% to 0.50 \%. \textbf{d} Average change in $R$ at all $V_{\text{G}}-V_{\text{D}}$ values, $\Delta R_{\text{avg}}$, versus $\Delta \varepsilon_{\text{mech}}$ in Device 2B (solid trace), compared to the theoretical calculations (dashed lines) at various $\theta$. \textbf{e} and \textbf{f} show the full theoretical calculations for Devices 2B and 2C, respectively.}
\label{fig.s7}
\end{figure}

Figures \ref{fig.s7}a-b show $R$ - ($V_{\text{G}}-V_{\text{D}}$) data for Device 2B (Fig. \ref{fig.s5}d) over the entire available range of $\varepsilon_{\text{mech}}$ ranging from 0.29 \% (dark blue trace) up to 0.87 \% (dark red trace), for forward and reverse mechanical sweeps, respectively. They show clearly that increasing $|A_{y}|$ increases $R$ continuously and reversibly. Based on the annealing data sequence in Fig. \ref{fig.s5}d, $\mu_{\text{contact}}$ is minimized in Device 2B which explains its very small transconductance. Figure \ref{fig.s7}c shows the  $R$ - ($V_{\text{G}}-V_{\text{D}}$) data in Device 2C (Fig. \ref{fig.s5}d) over a range of $\varepsilon_{\text{mech}}$ from 0.28 \% to 0.50 \%. The data at higher $\varepsilon_{\text{mech}}$ is not available for Device 2C, because the device failed after acquiring the data in \ref{fig.s7}c. The data for Device 2C was taken over a broader range of $V_{\text{G}}$ than for Device 2B, and show significantly more gate-dependence. We use the half-width half-maximum in Fig. \ref{fig.s7}c to estimate $n_{\text{rms}} \approx $ 4.0 $\times$ 10$^{11}$ cm$^{-2}$ in Device 2. This is considerably more than for Device 1, and it explains why its relative change in $R$ with $\varepsilon_{\text{mech}}$ is smaller than in Device 1. We note that such a large $n_{\text{rms}}$ also makes the theoretical calculations very weakly sensitive on the exact $\mu_{\text{contact}}$ value we use.

From Device 1 data, we know that the minimum  $\mu_{\text{contact}}$ achieved in our devices is $\approx$ 50 meV, and we use this value to model the data of Devices 2B and 2C. Figure \ref{fig.s7}d show $\Delta R_{\text{avg}}$ versus $\Delta \varepsilon_{\text{mech}}$ in Device 2B (solid trace) compared to the theoretical model (dashed lines) at using various $\theta$ values. This allows us to estimate $\theta \approx$ 10 $^{o}$ in Device 2. The full theoretical calculations for Devices 2B and 2C are in Fig.  \ref{fig.s7}e-f.

\section{Additional data on mechanically controlled quantum interferences}
\begin{figure}[hbtp]
  \centering
\includegraphics[scale=1]{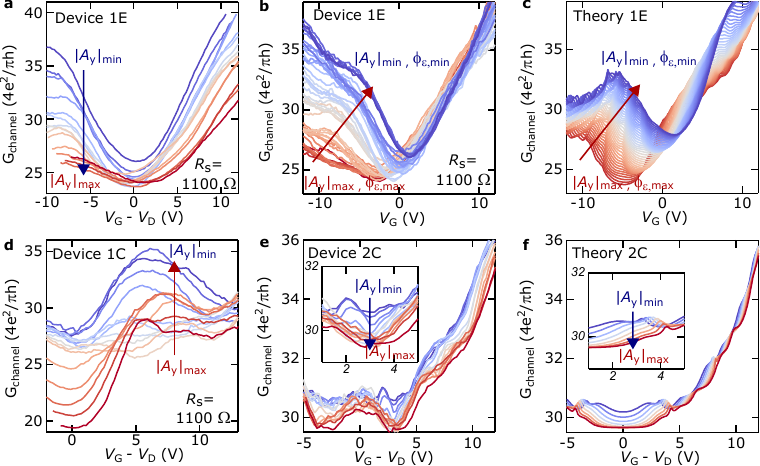}
\caption{Additional data on the mechanical control of quantum interferences.  \textbf{a} Forward mechanical sweep of $G_{\text{channel}}$ - ($V_{\text{G}}-V_{\text{D}}$) data versus $\varepsilon_{\text{mech}}$ in Device 1E for $\varepsilon_{\text{mech}}=$ 0 to 1.04 \%. \textbf{b} Raw reverse mechanical sweep in Device 1E as in Fig. 5a, without averaging traces three-by-three and without removing $\Delta V_{\text{D}}$. \textbf{c} Theoretical modeling of the data in \textbf{b}. \textbf{d} Reverse mechanical sweep of $G_{\text{channel}}$ - ($V_{\text{G}}-V_{\text{D}}$) data at $\varepsilon_{\text{mech}}=$ 1.04 \% (red) to 0 \% (blue) in Device 1C. \textbf{e} Forward mechanical sweep of $G_{\text{channel}}$ - ($V_{\text{G}}-V_{\text{D}}$) data at $\varepsilon_{\text{mech}}=$ 0.28 \% (blue) to 0.50\% (red) in Device 2C. \textbf{f} Theoretical modeling of the data in \textbf{e}.}
\label{fig.s8}
\end{figure}

Figure \ref{fig.s8} shows additional data from Devices 1 and 2, where quantum transport interferences are tuned mechanically, to support the data in Fig. 5 of the main text.
\BeginNoToc
\bibliography{supbib}
\EndNoToc